\documentclass[10pt]{iopart}

\pdfoutput=1 

\usepackage{graphicx}
\usepackage{float}
\usepackage[T1]{fontenc}
\usepackage{siunitx}
\DeclareSIUnit\year{yr}
\DeclareSIUnit\cts{cts}
\DeclareSIUnit\fwhm{FWHM}
\usepackage{xspace}
\usepackage{multirow}
\usepackage[cases]{cases}
\usepackage[switch]{lineno}
\usepackage[normalem]{ulem}
\usepackage[colorlinks=true]{hyperref}
\usepackage[all]{hypcap}
\usepackage[usenames, dvipsnames, table]{xcolor} 
\usepackage{placeins}
\usepackage[ignoreunlbld,nocites,norefs]{refcheck}
\usepackage{amssymb}
\renewcommand{\arraystretch}{1.25} 
\usepackage{comment}

\newcommand{\red}{\textcolor{red}}

\definecolor{mygreen}{rgb}{0.0, 0.5, 0.0}
\definecolor{myblue}{rgb}{0.0, 0.0, 0.5}

\newcommand{\0}{\ensuremath{0\nu\beta\!\beta}}
\newcommand{\2}{\ensuremath{2\nu\beta\!\beta}}
\newcommand{\bb}{\beta\!\beta}
\newcommand{\Q}{\ensuremath{Q_{\bb}}}
\newcommand{\mbb}{\ensuremath{\langle m_{\bb} \rangle}}
\newcommand{\T}{\ensuremath{T^{0\nu}_{1/2}}}
\newcommand{\isotope}[2]{$^{#2}$#1}
\newcommand{\geant}{Geant4}
\newcommand{\nest}{NEST}

\usepackage[draft, inline, nomargin, final]{fixme}
\fxsetup{theme=color,mode=multiuser}
\FXRegisterAuthor{mh}{anmh}{\color{red}MH}	
\FXRegisterAuthor{jb}{anjb}{\color{orange}JB}
\FXRegisterAuthor{lk}{anlk}{\color{myblue}LK} 
\FXRegisterAuthor{ss}{anss}{\color{blue}SS} 
\FXRegisterAuthor{ts}{ants}{\color{green}ts} 

\begin{document}
\title[nEXO: \0 search beyond $10^{28}$ year half-life sensitivity]{nEXO: Neutrinoless double beta decay search beyond $10^{28}$ year half-life sensitivity}


\author{G~Adhikari$^{1}$, 
S~Al~Kharusi$^{2}$, 
E~Angelico$^{3}$, 
G~Anton$^{4}$, 
I~J~Arnquist$^{5}$, 
I~Badhrees$^{6}$\footnote{Permanent at: King Abdulaziz City for Science and Technology, Riyadh, Saudi Arabia}, 
J~Bane$^{7}$, 
V~Belov$^{8}$, 
E~P~Bernard$^{9}$, 
T~Bhatta$^{10}$
% ...
}

\address{$^{1}$Physics Department, University of California, San Diego, CA 92093, USA}
\address{$^{2}$Physics Department, McGill University, Montr\'eal, Qu\'ebec H3A 2T8, Canada}
\address{$^{3}$Physics Department, Stanford University, Stanford, CA 94305, USA} \address{$^{4}$Erlangen Centre for Astroparticle Physics (ECAP), Friedrich-Alexander University Erlangen-N\"urnberg, Erlangen 91058, Germany} \address{$^{5}$Pacific Northwest National Laboratory, Richland, WA 99352 USA} \address{$^{6}$Department of Physics, Carleton University, Ottawa, Ontario, K1S 5B6, Canada} \address{$^{7}$Amherst Center for Fundamental Interactions and Physics Department, University of Massachusetts, Amherst, MA 01003, USA} \address{$^{8}$Institute for Theoretical and Experimental Physics named by A. I. Alikhanov of National Research Center Kurchatov Institute, Moscow, 117218 Russia} \address{$^{9}$Lawrence Livermore National Laboratory, Livermore, CA 94550, USA} \address{$^{10}$Department of Physics and Astronomy, University of Kentucky, Lexington, KY 40506, USA}
% ...

\ead{\mailto{sangiorgio1@llnl.gov} and \mailto{caio.licciardi@snolab.ca}}


\date{\today{}}

\begin{abstract}
The nEXO neutrinoless double beta (\0) decay experiment is designed to use a time projection chamber and 5000~kg of isotopically enriched liquid xenon to search for the decay in \isotope{Xe}{136}. Progress in the detector design, paired with higher fidelity in its simulation and an advanced data analysis, based on the one used for the final results of EXO-200, produce a sensitivity prediction that exceeds the half-life of $10^{28}$ years.  Specifically, improvements have been made in the understanding of production of scintillation photons and charge as well as of their transport and reconstruction in the detector. The more detailed knowledge of the detector construction has been paired with more assays for trace radioactivity in different materials. In particular, the use of custom electroformed copper is now incorporated in the design, leading to a substantial reduction in backgrounds from the intrinsic radioactivity of detector materials. 
Furthermore, a number of assumptions from previous sensitivity projections have gained further support from interim work validating the nEXO experiment concept. 
Together these improvements and updates suggest that the nEXO experiment will reach a half-life sensitivity of $1.35\times 10^{28}$~yr at 90\% confidence level in 10 years of data taking, covering the parameter space associated with the inverted neutrino mass ordering, along with a significant portion of the parameter space for the normal ordering scenario, for almost all nuclear matrix elements.  The effects of backgrounds deviating from the nominal values used for the projections are also illustrated, concluding that the nEXO design is robust against a number of imperfections of the model.

\end{abstract}



\maketitle
\ioptwocol
\flushbottom

\section{Introduction}

Neutrinoless double beta (\0) decay is an 
undiscovered hypothetical process in which a nucleus with mass $A$ 
and charge $Z$ undergoes the decay $(A, Z) \rightarrow (A, Z + 2) + 2\,e^{-}$ 
without emission of neutrinos~\cite{PhysRev.56.1184}. 
The search for \0 decay is considered one of the most sensitive  
tests of the Majorana nature of neutrinos~\cite{PhysRev.56.1184,doi:10.1142/S0218301311020186}.  
The observation of \0 decay would open a portal to new physics beyond the Standard Model by 
providing the first direct evidence for the violation of lepton number conservation, 
with possible implications for our understanding of the observed baryon asymmetry in the universe~\cite{Fukugita:1986hr,Rubakov_1996,PhysRevD.98.055029, leptogenesis}. 
Under the standard assumption that the decay is mediated by a light Majorana neutrino, the discovery of the decay would help constrain the absolute neutrino mass scale since  
the experimentally observable \0 decay half-life ($\T$) is inversely proportional 
to the square of the effective Majorana neutrino mass $\mbb$~\cite{Avignone:2007fu}. 

\isotope{Xe}{136}, one of the  even-even nuclei with energetically forbidden $\beta$ decay~\cite{Zyla:2020zbs}, is an attractive nuclide for a \0 decay search due to several reasons. 
A \0 signal would be a collection of events with topology consistent with double-$\beta$ decays, homogeneously distributed in the detector, uncorrelated in time with other interactions, and with an average energy consistent with the known $Q$-value of the decay. 
Since the \isotope{Xe}{136} double-$\beta$ decay energy is \Q$=2458.10\pm0.31$~keV~\cite{PhysRevLett.98.053003,McCowan:2010zz}, it is large compared to $Q$ values of many natural radioactive decays.
As a noble gas, 
it can be made largely free of radioactive contamination 
through repeated purification. 
In addition, isotopic separation of \isotope{Xe}{136} from the 8.9\% natural concentration is relatively inexpensive and 
has already been carried out at the tonne scale~\cite{zencollaboration2021nylon}.    While uncertainties still exist in the estimates of nuclear matrix elements (NMEs)~\cite{Caurier2008,Rodriguez2010,Suhonen2010,Simkovic2013,Mustonen2013,Meroni2013,Vaquero2013,Engel2014,Hyvaerinen2015,Barea2015,Yao2015,Horoi2016,Menendez2018,Simkovic2018,Jokiniemi2018,Fang2018,Deppisch2020}, that govern the \0 decay process, the product of the squared NME and phase space factor is expected to be favorable for \isotope{Xe}{136} in comparison to some other candidate nuclides, such that the sensitivity per unit mass to the underlying physics places it among the top contenders for large scale detectors \cite{Biller:2021bqx}.  

The success of the 100-kg scale EXO-200 experiment~\cite{PhysRevC.89.015502,PhysRevLett.109.032505,EXO_Nature,PhysRevLett.120.072701,PhysRevLett.123.161802} validates using large liquid xenon (LXe) time projection chambers (TPCs) for the search of \0 decay. This technique provides a large homogeneous mass of the required isotope with three-dimensional event vertex reconstruction and well understood energy resolution.
EXO-200 was the first experiment to observe two-neutrino double-$\beta$ decay (\2) in \isotope{Xe}{136}~\cite{PhysRevLett.107.212501}. EXO-200 then precisely measured its half-life of $2.165\pm0.016\text{ (stat)}\pm0.059\text{ (syst)}\times10^{21}$~yr~\cite{PhysRevC.89.015502}, 
and set stringent constraints on the \0 decay process, limiting the half life to $\T > 3.5\times10^{25}$~yr 
at 90\% confidence level (CL), with corresponding sensitivity of $5.0\times10^{25}$~yr 
using an exposure of 234.1~kg$\cdot$yr~\cite{PhysRevLett.123.161802}.
Compared to other candidate systems, a LXe TPC offers the further advantage that the active material can be recirculated, and hence its purity can be improved with time. 

The advantages of the LXe TPC have a greater impact on the experiment 
as the detector size increases, and the tonne-scale nEXO detector has been designed to exploit these advantages, reaching a $\T$ sensitivity beyond $10^{28}$~yr with 5000~kg of xenon enriched to 90\% \isotope{Xe}{136}~\cite{nEXOpCDR}.  Following its predecessor, nEXO will simultaneously measure the signals from both the scintillation light and from the drifting ionization, combining the two to obtain three-dimensional ``images'' of the energy depositions.  Event reconstruction topology forms the basis of discrimination between double-$\beta$ decay events and background events which are mainly gamma-induced signals.  In a LXe medium the former generally have energy concentrated in one spatial location, or site, with the events distributed uniformly throughout the detector volume, while the latter result in energy deposited in multiple sites, with event rate decreasing exponentially toward the center of the detector due to attenuation in the dense LXe.  Internal $\alpha$ backgrounds are tagged and eliminated based on their larger track ionization density which results in a large scintillation-to-ionization ratio compared to $\beta$-like events.  The energy measurement makes use of the anticorrelation~\cite{PhysRevB.20.3486} between scintillation and ionization signals by independent measurements of these quantities~\cite{PhysRevB.68.054201}, obtaining a resolution that allows the rejection of the \2 decay background to negligible levels for the sensitivities of interest here.

In an initial analysis~\cite{PhysRevC.97.065503}, nEXO was shown to reach a \0 sensitivity of $9.2\times10^{27}$~yr when using a multi-parameter analysis similar to that featured in early analyses of EXO-200 data~\cite{EXO_Nature,PhysRevLett.120.072701}. 
This report presents an improved sensitivity estimate, applying a more realistic model that reflects advances in the design and 
understanding of the nEXO performance obtained through ongoing R\&D~\cite{nEXOpCDR,PhysRevC.97.065503,Li_2019,Nakarmi_2020,Jamil:2018tkx,Gallina:2019fxt,Ostrovskiy:2015oja,Wagenpfeil:2019qri,stiegler2020event,NJOYA2020163965}. 
It also includes improved knowledge of the ionization and scintillation production in LXe at MeV energies~\cite{PhysRevC.101.065501} and increased sensitivity in material assay measurements. 
We also incorporate a deep neural network (DNN) developed to discriminate between signal and background events, following the work done by EXO-200 for their \0 decay search using their complete dataset~\cite{PhysRevLett.123.161802}. 
The refined geometrical design of nEXO and its implementation in Monte Carlo (MC) simulations are introduced in section~\ref{sec:geometry}.
Section~\ref{sec:simulation} describes the MC procedure to simulate and reconstruct  
nEXO events as well as the development of a \0 DNN discriminator. 
The expected background budget is then discussed in section~\ref{sec:materials}. 
The discovery potential and half-life sensitivity of nEXO to \0 decay are presented in Section~\ref{sec:sensitivity} and final results are summarized in Section~\ref{sec:conclusion}.

\section{Refined Geometric Design}\label{sec:geometry}

\begin{figure*}[tbp]
    \centering
    \includegraphics[width=0.45\textwidth]{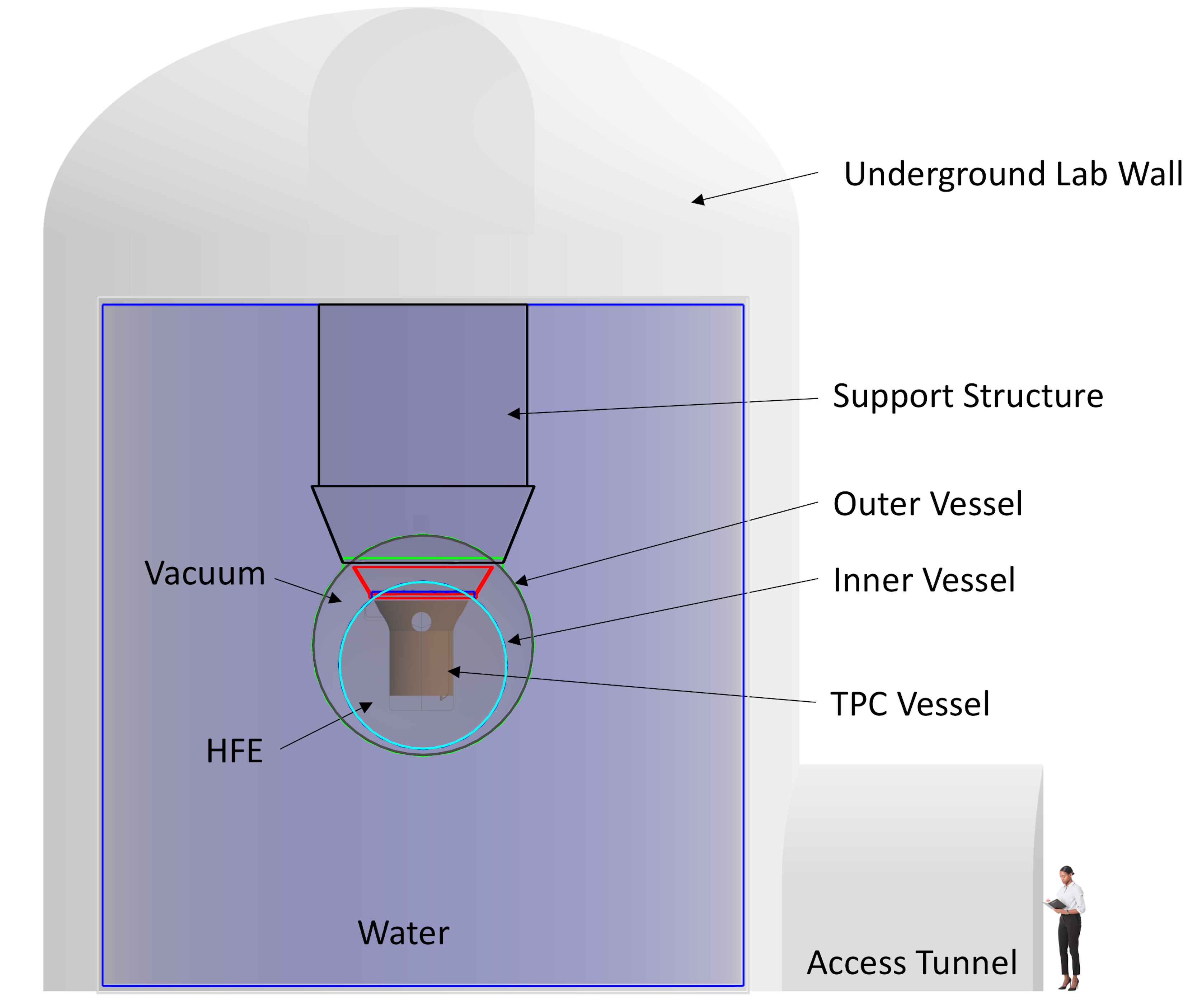}
    \hspace{5mm}
    \includegraphics[width=0.45\textwidth]{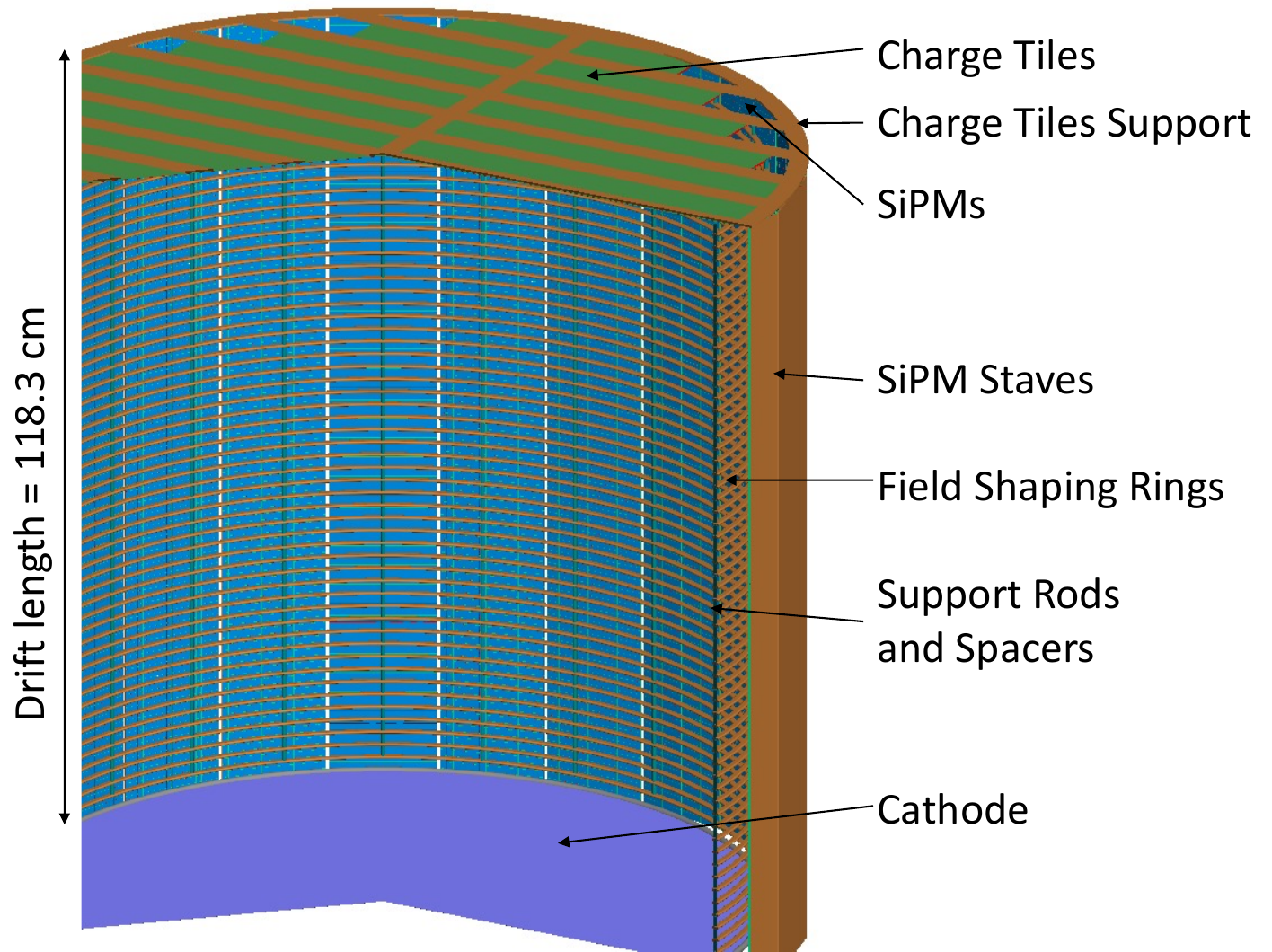}
    \caption{Visualization of the \geant{} geometry implementation for MC simulations of the nEXO detector. The drawing on the left shows a cross-section view of the large components external to the TPC vessel, including the OD's water and SNOLAB's Cryopit wall. The model on the right shows a close up view of the main components inside the TPC vessel.}
    \label{fig:g4_geometry}
\end{figure*}

As shown on the left side of Figure~\ref{fig:g4_geometry}, the design of the nEXO detector~\cite{nEXOpCDR} 
consists of a TPC vessel filled with enriched LXe and surrounded by 
${\sim33,000}$~kg of HFE-7000~\cite{HFE-7000}, which serves as both thermal bath and radiation shield. A vacuum layer between an inner vessel (IV) and an outer vessel (OV) of the cryostat provides thermal insulation from an active water shield, 
referred to as the outer detector (OD). Updated dimensions in the TPC vessel and drift region represent the most significant difference in the current geometry with respect to that employed in  previous studies~\cite{nEXOpCDR,PhysRevC.97.065503}. 

In the current design, the TPC vessel is a right copper cylinder with both inner height and diameter  
equal to 127.7~cm. This represents a small reduction in size compared to~\cite{PhysRevC.97.065503}, and corresponds to a total mass of 4811~kg of contained LXe when accounting for 
the volume displaced by internal components and using a density of 3.057 g/cm$^3$ for the enriched xenon. 
The total drift length between the cathode and the anode is 118.3~cm as indicated on the right side of Figure~\ref{fig:g4_geometry}. 
At the anode, located on the top end of the TPC, charge is collected by 0.6~cm pitch and 9.6~cm long crossed electrode strips arranged as a large array of 9.6$\times$9.6~cm$^2$ interconnected readout tiles filling the top face of the TPC. A detailed description of the anode is provided in Ref.~\cite{nEXOpCDR,Jewell:2017dzi}. The advantage in background discrimination from employing a smaller pitch than that between strips is outweighed by the benefits of reducing the number of digitizing channels by a factor of two, resulting in a significant reduction of cost and mass of readout cabling~\cite{Li_2019}.

Electrostatic modeling of the TPC was performed to ensure that the electric field remains below a maximum allowed strength everywhere to prevent breakdown. 
To achieve 400~V/cm electric field within the field cage region, we require gaps between the vessel end  
and cathode of 7.5~cm, as well as 7.2~cm spacing between the barrel surface of the xenon vessel and the field shaping rings (FSRs). 
In this volume, VUV-sensitive silicon photomultipliers (SiPMs) positioned in a ``barrel'' configuration detect LXe scintillation light.  
A smaller 1.9~cm gap between the anode and the top of the vessel is provided to accommodate electronics and cabling with sufficient mechanical clearance. 
The inner radius of the FSRs is 56.7~cm, yielding a fully active TPC LXe mass of 3648~kg. The remaining volume, where only the scintillation light but no ionization from LXe interactions is detected, is referred to as the ``skin'' Xe region~\cite{stiegler2020event}. 

\begin{table}[tbp]
\centering
\begin{tabular}{ll} 
\hline
\textbf{Description} & \textbf{Value} \\ 
\hline 
Liquid xenon mass in vessel & 4811~kg \\
Liquid xenon mass in drift region & 3648~kg \\
Fiducial xenon mass & 3281~kg \\ 
TPC drift height & 118.3~cm \\
TPC drift diameter & 113.3~cm \\
TPC vessel height & 127.7~cm\\
TPC vessel diameter & 127.7~cm\\
Inner vessel diameter* & 338~cm\\
Outer vessel diameter* & 446~cm \\
Water Tank height & 13.3~m \\
Water Tank diameter & 12.3~m\\

\hline
\end{tabular}
\caption{Main dimensions of the nEXO geometry. The two cryostat dimensions marked with a *  indicate values that were not changed from the previous analysis~\cite{PhysRevC.97.065503}.} 
\label{tab:key_params}
\end{table}

Dimensional specifications for the SiPMs and charge readout tiles,  
as well as for their support structures and electronics 
are adjusted to the slightly smaller vessel size, along with other minor design adjustments. Table~\ref{tab:key_params} summarizes the corresponding primary dimensions. The geometry implemented in MC simulations, built using \geant~v10.4~\cite{AGOSTINELLI2003250,1610988,ALLISON2016186} and  
described in more detail in Ref.~\cite{PhysRevC.97.065503},   
has been updated to match these dimensions as specified in the latest engineering model. As described below, we have also added components outside the TPC vessel that were not included in the previous work. The result is the detailed representative model of the experiment shown in Figure~\ref{fig:g4_geometry}. A full list of the simulated components accounted for in the nEXO background model is discussed in Section~\ref{sec:materials}. 

The MC geometry now includes the high voltage (HV) feedthrough and two tubes within the HFE-7000 volume that are designed to bring calibration sources near the outside of the TPC vessel. The support structure for both IV and OV have been implemented along with realistic attachments and feedthroughs, and with thicknesses matching recent engineering estimates. In particular, we added additional 0.2~cm and 0.3~cm thick titanium internal liners to both the IV and OV, respectively. They provide the winding mandrel for the carbon-fiber composite (CFC) employed in these vessels. For the IV, the liner also doubles as an adsorption barrier for the HFE-7000. In the model, the liners are made of titanium but other materials are also under consideration. All of these components are modeled as both passive radioactive background shields and background sources through dedicated simulations of decays in the materials. Background contributions from electronics cables, running along the vessel structures outside of the TPC vessel, were found to be negligible, and hence are not included in the simulation.  

The OD geometry presented here has been designed for installation in the Cryopit at SNOLAB. The OD consists of a water tank enclosing the OV that serves two purposes: 
to shield against external radiation, and to tag the passage of nearby muons, which allows for vetoing of cosmogenically-induced events in the TPC. Our MC simulations indicate that a minimum diameter of 11~m and a height of 12~m results in enough water shielding to sufficiently reduce background contributions from the cavern rock and concrete, and to produce nearly maximal muon tagging efficiency when photomultiplier tubes (PMTs) are mounted on the tank wall and used to read out the Cherenkov light in the water. The simulated nEXO geometry contains a water tank, 12.3~m in diameter and 13.3~m in height, which is compatible with the current dimensions of the Cryopit.

\section{Event Simulation and Reconstruction}\label{sec:simulation}

This work improves the fidelity of the expected detector response to scintillation photons and direct ionization produced in LXe. Electronic waveforms produced by the charge signals are simulated in detail, as well as the reconstructed event energy and topology. The simulation takes into account generation of charge carriers and scintillation photons, their propagation in LXe, and electronic signals captured by readout. We then use the simulated signals and reconstructed energies to determine the event parameters that are input to the \0 sensitivity analysis.

\subsection{Ionization/scintillation anticorrelation in LXe}

A custom version of the Noble Element Simulation Technique (\nest{})~\cite{Szydagis:2011tk} calculates the number of electrons ($N_{Q}$) and scintillation photons ($N_{P}$) produced in each simulated energy deposit in the LXe. These values are stored in a list of hits, along with their generated positions. 

\nest{} version 2.0.1~\cite{nest2.0.1} was modified to correct interface issues with \geant{}. Furthermore, two adjustments were made in the model to attain better agreement with EXO-200 results~\cite{PhysRevC.101.065501}. First, \nest{} 2.0.1 included an excess noise correlation in the total number of created electrons and photons (i.e. an inflated Fano factor), which lacked physical motivation and did not agree with the noise model determined by EXO-200. It was therefore removed for this analysis, later versions of \nest{} incorporated this change. Second,  the \nest{} photoelectric model is replaced by the NEST beta/Compton model which is used for all recoil electron processes. This conforms with EXO-200 results which observed no significant difference in yields between $\beta$ decays and photoelectric interactions at energies relevant for nEXO. 

While light and charge yields in the fully active TPC region are only mildly affected by these changes, we noticed that regions with high electric field, such as in the skin Xe, are substantially impacted. The electric field used in the modified \nest{} calculations is considered to be uniform in six different regions of the LXe as indicated in Table~\ref{tab:voltage_regions}. Edge regions, where the electric field is not expected to be uniform, are either discarded in the analysis or represent a negligible fraction of the total volume, so that the uniform approximation does not affect the final results.  

Some lack of fidelity is expected in the regions outside of the field cage, because the currently available data interpolated in \nest{} does not reach sufficiently high field values. The skin region has electric fields that vary from 0~V/cm near the anode to over 8000~V/cm near the cathode. While EXO-200 data is used to establish light and charge yields in the 400~V/cm field in the TPC, these data only extend to 600~V/cm and do not reach the higher fields present in the skin region. For this study, the light and charge yields were fixed to their values at 600~V/cm for all greater values of the field. This approach was chosen in order to be conservative, as it leads to slightly higher backgrounds compared to models in which the yields continue to decrease with electric field beyond 600~V/cm. Future work will evaluate the yield dependence at higher field values, which has been reported to be small but nonzero \cite{PhysRevB.20.3486, Doke:1988rq}. This could slightly reduce backgrounds compared to this study's model.

\begin{table}[tbp]
    \centering
\begin{tabular}{ccc}
 \hline
 & \multicolumn{2}{c}{Radial Position}                                                                   \\
& Inside FSR & Outside FSR \\ 
\hline                                       
Above anode                                                                      & \multicolumn{1}{c}{0 V/cm}      & 0  V/cm                                                               \\ \hline
\begin{tabular}[c]{@{}c@{}}Between anode\\ and cathode\end{tabular}      & \multicolumn{1}{c}{400 V/cm}       & \begin{tabular}[c]{@{}c@{}}Linear from \\ 0 to 8084 V/cm\end{tabular} \\ \hline
Below cathode                                                                & \multicolumn{1}{c}{6667 V/cm}     & 8084  V/cm      \\ \hline                                                       
\end{tabular}
    \caption{Electric field settings used in \nest{}'s calculations of charge and light yield in the LXe for inside and outside the field shaping rings (FSRs).}
    \label{tab:voltage_regions}
\end{table}

\subsection{Charge signal detection}

The simulation of the charge waveforms follows from the successful procedure in EXO-200 and has been validated in recent R\&D data. It starts by modeling the drift of each ionization hit from its initial position to the charge readout strips at the anode. The electrons are drifted at a fixed velocity of 0.171~cm/$\mu$s, 
appropriate for the nominal electric field of 400~V/cm. The corresponding transverse and longitudinal diffusion coefficients, measured in Refs.~\cite{PhysRevC.95.025502,NJOYA2020163965}, are used to appropriately broaden the charge distribution in the $x$, $y$, and $z$ dimensions. 
Diffusion is applied before binning the electrons into cubic voxels used to calculate the induced charge on each electrode as a function of time \cite{Li_2019}. Charge attenuation depends on the drift distance and assumes an electron lifetime of $\tau_e=10$~ms. The waveforms are simulated for all strips with charge collection 
as well as for channels within a distance of 1.8~cm to a charge collection strip, prior to adding electronics noise.

\begin{figure}[tbp]
    \centering
    \includegraphics[draft=false,width=\columnwidth]{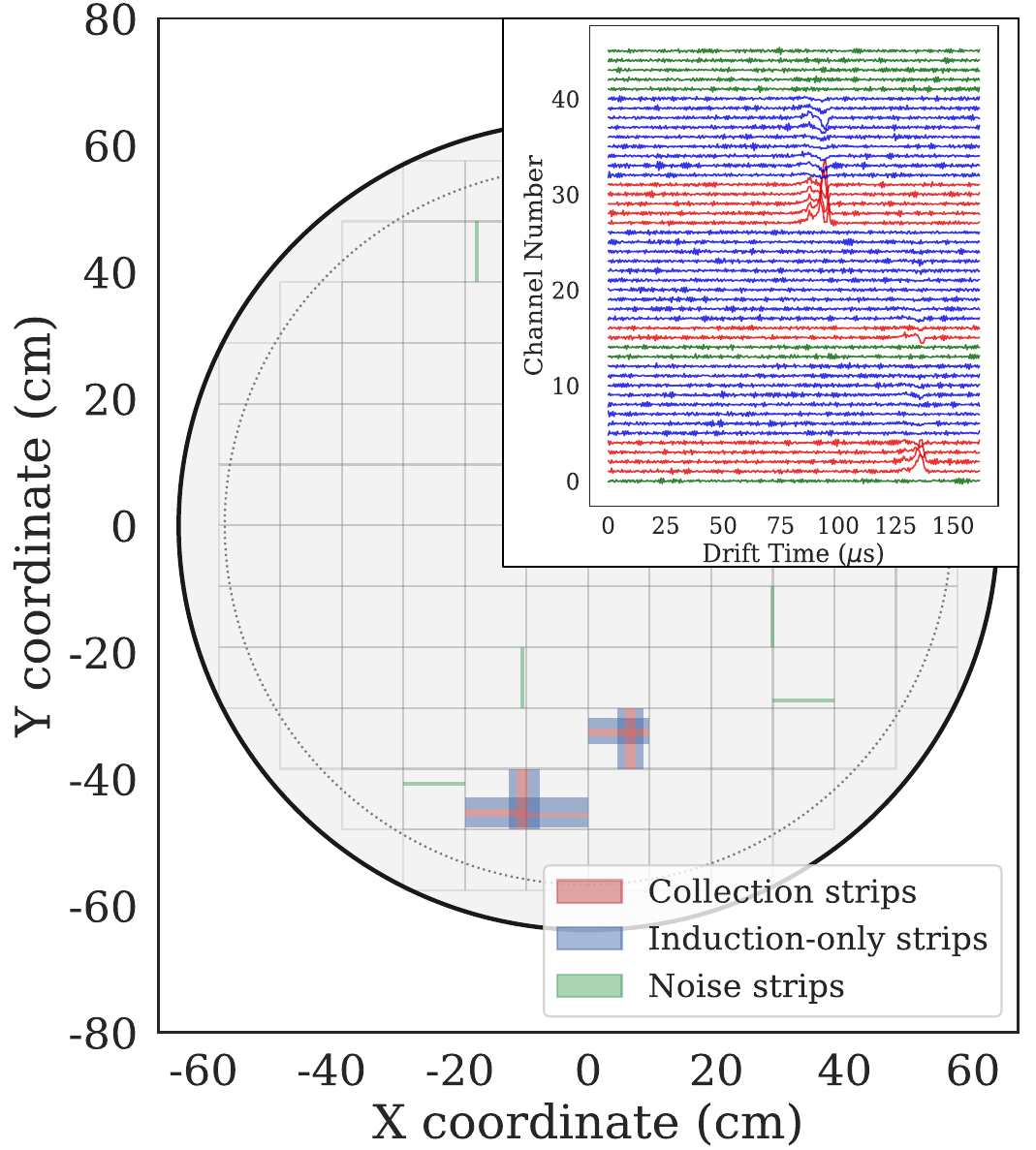}
    \caption{Top view of the charge readout tiles (squares outlined by dotted lines) and current waveform display (inset) showing a multi-site event with energy deposits in three tiles. The red strips and waveforms indicate charge collection, blue strips and waveforms represent induced charge without collection, and the green waveforms only contain noise that triggers the reconstruction algorithm.}
    \label{fig:charge_waveforms}
\end{figure}

The frequency spectrum of the noise in the charge electronics, as determined in recent R\&D with prototype ASICs, is sampled and added to the waveform after transformation into the time domain. The resulting charge waveform is converted into a time series of the current. A Bessel filter with a cut-off frequency of 300~kHz is then applied to reproduce the analog anti-aliasing filter in the electronics, 
and the result is down-sampled to the 2~MHz sampling rate of the nEXO electronics design. The simulated current noise is scaled to have an equivalent RMS value as deduced from simulations of the front-end electronics at the nominal design specifications.
More details about the waveform simulation can be found in Ref.~\cite{Li_2019}. Figure~\ref{fig:charge_waveforms} shows the simulated waveforms 
for an example multi-site event with all energy deposits localized in three charge tiles. 

A fiducial volume (FV) selection is applied to the position of the outermost energy deposit of each event in the TPC, reconstructed from information of the charge collected in each strip. To reconstruct the $z$ position, the event start time is subtracted from the charge waveform peak timing and the difference is converted into a distance using the fixed electron drift velocity. The outermost energy deposit is found by scanning along the $x$ and $y$ coordinates of strips which collect charge at matching $z$ positions to find the outermost position. Each event is required to have at least one strip hit in both $x$ and $y$, reducing the \0 detection efficiency by 3.6\%. The spatial resolution obtained for such deposits is better than the 0.6~cm pitch of the charge-sensing strips in all three dimensions. 
The relative distance between a deposit and the nearest surface (\emph{i.e.} the FSRs, the cathode, or the anode planes) is referred to as the event's \textit{standoff distance}. Electric field simulations set a minimum standoff distance of 2~cm required for an event to be within the uniform field, 
which is also a sufficient condition to reject $\beta$ particles from decays on the surface of the nearest components to the FV. With this choice of FV, the LXe mass included in the \0 analysis is 3281~kg.  

The number of electrons detected in each event is calculated using information from the charge waveforms in a simplified algorithm that retains only the main features of this process, described as follows. The true charge collected on all strips is summed up, corrected for the electron lifetime attenuation, and Gaussian noise with $\sigma_Q = 1130$~e$^{-}$ is added, resulting in the corrected charge of the event ($\tilde{N}_Q$). The choice for the electronic noise is an average value obtained from a MC study using simulated waveforms from \0 decays. In this study, a trapezoidal filter with optimal integration time was employed to improve the charge energy resolution, which was then used to deduce the equivalent average noise corresponding to this resolution, $\sigma_Q$. 

\subsection{Scintillation energy}

In order to simulate the number of scintillation photons detected for each event, 
the light collection efficiency $\varepsilon_\ell(r,z)$ was calculated as a function of the radial ($r$) and longitudinal ($z$) initial position inside the detector. 
$\varepsilon_\ell(r,z)$ is the product of the photon transport efficiency (PTE), that depends on the initial position, and the photon detection efficiency (PDE) of the SiPMs. In Refs.~\cite{nEXOpCDR,PhysRevC.97.065503}, light propagation in the nEXO detector has been studied using \geant{}. Photons were generated by point-like isotropic sources in the LXe and tracked through the TPC. Optical photon propagation in \geant{} is computationally expensive. An alternative approach was therefore employed using Chroma running on CUDA-enabled GPUs~\cite{Seibert2011FastOM,chroma}, making light propagation simulations up to 300 times faster. Geometries were directly imported into Chroma from standard CAD software allowing an exact remake of the detector geometry without much simplifications. 

The optical parameters that are used as input for the simulations 
have been refined, leveraging recent measurements, and are summarized in Table~\ref{tab:optical_parameters}. The LXe index of refraction was calculated using the approach in Ref.~\cite{Hitachi} which is in agreement with measurements \cite{SOLOVOV2004462}. Measurements of the absorption ($\lambda_\mathrm{abs}$) and Rayleigh scattering ($\lambda_\mathrm{scat}$) lengths at $175$~nm wavelength~\cite{FUJII2015293}, with sufficient sensitivity for the light propagation model, are difficult to obtain.  Other large LXe detectors assume $\lambda_\mathrm{abs}$ to be the values 30~m \cite{Mount:2017qzi} and 50~m \cite{Aprile:2015uzo}.  Here we conservatively assume $\lambda_\mathrm{abs}=20$~m that should be easily achieved given the much higher LXe purity imposed by the electron drift requirements and the extrapolation from Ref.~\cite{Baldini:2004ph}.
Likewise, we conservatively assume $\lambda_\mathrm{scat} = 30$~cm, as supported by Refs.~\cite{SOLOVOV2004462, Virdee:1991rv, CHEPEL1994500}.

\begin{table}[t]
\centering
\begin{tabular}{l|c}
\hline
\textbf{Parameter} & \textbf{Value} \\ 
\hline

LXe refractive index & 1.69 \\
LXe $\lambda_\mathrm{abs}$ & \SI{20}{\meter} \\
LXe $\lambda_\mathrm{scat}$ & \SI{30}{\centi\meter} \\ 
$R_S$(TPC Vessel)  & \SI{0}{\percent}  \\
$R_S$(Cathode)  & \SI{80}{\percent}  \\
$R_S$(Anode) & \SI{20}{\percent} \\
$R_S$(Field Rings) & \SI{80}{\percent}  \\
$R_S$(Support Rods) & \SI{0}{\percent} \\
$R_S$(SiPM Staves) & \SI{0}{\percent}  \\
$R_S$(SiPMs)$[15^\circ < \theta < 45^\circ]$ & \SI{25}{\percent} - \SI{28}{\percent} \cite{Nakarmi_2020}\\

\hline
\end{tabular}
\caption{Optical parameters and reflectivity values used in the LXe optical simulations.}
\label{tab:optical_parameters}
\end{table}

Two surface models are implemented and used for optical simulations depending on the component. 
A basic model is used for all non-detecting surfaces, 
including the FSRs, cathode, anode, FSR support rods, SiPM staves and TPC vessel. 
At the surface of these components, the photon interaction is determined by their coefficients of specular reflectivity ($R_S$) and absorption, while the diffuse reflectivity is always set to zero (the latter being a conservative assumption). In order to increase light collection, the cathode and FSRs components feature a reflective $\text{Al}+\text{MgF}_2$ coating with a conservative 80\% specular reflectivity~\cite{vuv_mirror2}. The angular dependence of such coatings has been studied using FreeSnell~\cite{freesnell}, a thin film optical simulator, and was estimated to be only different by $\sim \SI{10}{\percent}$ at large angles of incidence compared to a flat \SI{80}{\percent} specular reflectivity. For the anode, composed of the gold-plated charge readout tiles, a 20\% specular reflectivity was adopted based on measurements of the intrinsic reflectivity of gold at 175~nm~\cite{doi:10.1063/1.3243762}.

\begin{figure}[t]
    \centering
    \includegraphics[width=\columnwidth]{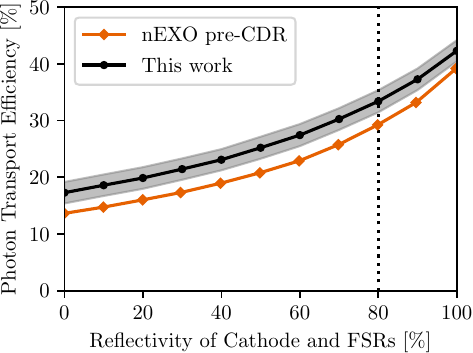}
    \caption{Spatial average PTE inside the fiducial volume as a function of the specular reflectivity of the cathode and the FSRs. The current light simulation results with updated optical parameters 
    (black circles) are compared against results from previous simulations 
    (orange diamonds)~\cite{nEXOpCDR}. The grey band indicates a systematic error due to the choice of SiPMs and light simulator. The reflectivity value used in the simulations for this work is indicated by the vertical dotted line.}
    \label{fig:pte}
\end{figure}

\begin{figure}[t]
    \centering
    \includegraphics[width=\columnwidth]{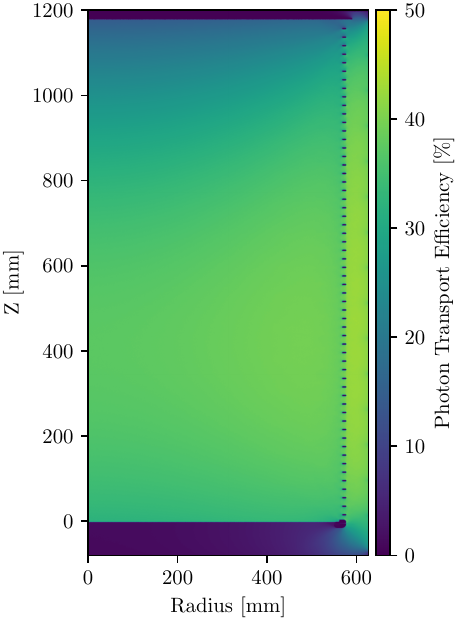}
    \caption{Light map of the nEXO detector showing the  probability of photons reaching the avalanche region of the SiPMs, \emph{i.e.}, the PTE as function of the radial, $r$, and longitudinal, $z$, positions in the detector. TPC components such as the FSRs, the anode and cathode are clearly visible due to their high opacity.}
    \label{fig:lightmap}
\end{figure}

At the detecting SiPM surface, the surface model incorporates recent measurements of the reflectivity and relative PDE as a function of the incidence angle in LXe~\cite{Nakarmi_2020,Wagenpfeil:2021ynz}. The nEXO collaboration is undertaking an extensive campaign for characterizing SiPMs from different vendors to determine the optimal candidate that meets stringent requirements in terms of PDE, correlated avalanche rate, dark noise rate, capacitance per unit area and intrinsic radiopurity~\cite{nEXOpCDR,Ostrovskiy:2015oja, Jamil:2018tkx, Gallina:2019fxt}. 
In particular, recent results indicate that both the FBK VUV-HD1~\cite{Jamil:2018tkx} and Hamamatsu VUV4~\cite{Gallina:2019fxt} satisfy such requirements. In order to evaluate the impact of using these different SiPMs, 
full light propagation simulations were run with each, as well as with different sets of reflectivity data. For each set of SiPM candidate and measured reflectivity data the average PTE within the FV was calculated as a function of reflectivity of the cathode and FSRs. Figure~\ref{fig:pte} shows the average over all these simulations. The gray band is an error calculated by adding the standard deviation across simulations with different SiPMs and the difference between Chroma and \geant{} of 1.8\% in quadrature. The expected average PTE for nEXO in the current design with 80\% reflectivity on the cathode and FSRs is \SI[separate-uncertainty = true]{33.3(20)}{\percent}. For the MC simulations, we use a conservative PTE lightmap, shown in Figure~\ref{fig:lightmap}, produced using \SI{5E11}{} photons and the Hamamatsu VUV4 SiPM parameters, which presents the lowest $\mathrm{PDE}/(1-R)$ value of 18.6\% as corrected for the reflectivity at near normal incidence in vacuum. This is done to prevent double-counting the effect of reflectivity when combining the PDE and PTE to estimate the total light collection efficiency $\varepsilon_\ell(r,z)$. While the PTE  is observed to change by about a factor of two between the anode and the center of the LXe volume, the absolute difference in the post-correction energy resolution of \0 events between these two regions was found to be only $\sim 0.1\%$. 

The corresponding $\varepsilon_\ell(r,z)$ from Figure~\ref{fig:lightmap} 
is applied to each scintillation hit ($\varepsilon_\ell^\text{hit}$) in an event, and  a binomial distribution with $N_P$ trials and $\varepsilon_\ell^\text{hit}$ probability is used to determine the number of detected photons: 
$N_{P}^\text{hit} \sim B(N_{P},\,\varepsilon_\ell^\text{hit})$. The contribution from correlated avalanches, in which a photon-induced avalanche in one cell of a SiPM can induce an avalanche in a neighboring cell, is modelled by a Poisson distribution: $N_{\Lambda}^\text{hit} \sim \text{Pois}(\Lambda \cdot N_{P}^\text{hit})$, where $\Lambda = 0.2$ is the total number of correlated avalanches per avalanche in a time window of 1~$\mu$s, conforming with recent measurements~\cite{Jamil:2018tkx,Gallina:2019fxt}. The same measurements also show that the dark count rate and other electronics noise can be completely neglected.
Therefore, the total number of scintillation photons triggering an avalanche 
in each event ($N_{P}^\text{evt}$) is the sum of $N_{P}^\text{hit}$ and $N_{\Lambda}^\text{hit}$ for all hits. The number of reconstructed photons ($\tilde{N}_P$) in each event is obtained by the correction $\tilde{N}_P = N_{P}^\text{evt}/\varepsilon_\ell^\text{evt}/(1+\Lambda)$, where $\varepsilon_\ell^\text{evt}$ is the energy-weighted average of all $\varepsilon_\ell^\text{hit}$. The position and energy information in this correction are retrieved from the scintillation hits generated by NEST. 
An additional systematic error due to imperfect knowledge of $\varepsilon_\ell^\text{evt}$ has been added and is assumed to be a \SI{0.5}{\percent} relative error. Based on our energy resolution model this is the value below which its contribution to the rotated energy resolution can be considered subdominant. Additional work to better understand position-dependent variations and the resulting requirements for the detector calibration will be part of future studies.

\subsection{Energy resolution}

\begin{figure}[t]
    \centering
    \includegraphics[width=\columnwidth]{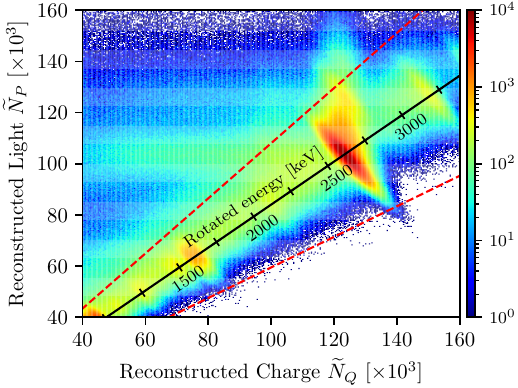}
    \caption{Reconstructed light and charge signals of all events in the FV from simulated \isotope{Th}{232} decays in the TPC vessel. The dashed lines indicate the location of the diagonal cut, described in the text. The rotated axis indicates the scale of the rotated  energy.}
    \label{fig:charge_light_cut}
\end{figure}

Figure~\ref{fig:charge_light_cut} shows the spectrum of the reconstructed number of generated photons, $\tilde{N}_P$, versus that of the electrons, $\tilde{N}_Q$, for simulated \isotope{Th}{232} decays in the TPC vessel. While the well-known anticorrelation between light and charge signals is clearly visible for the $\gamma$ lines, an excess of events with abnormally large light-to-charge ratio is present.
This effect is discussed in detail in Ref.~\cite{stiegler2020event}, and is caused by events that deposit some energy in the skin Xe. These events can be removed from the analysis using a diagonal cut, shown by the dashed lines, since the distribution of light-to-charge ratio is expected to be approximately a Gaussian around the mean ratio. Using simulated $\gamma$-rays from \isotope{Tl}{208} and \isotope{K}{40} decays with a single energy deposit inside the field cage, this cut was designed to keep 99\% of events, \emph{i.e.}, within 2.57$\sigma$ from the mean of the distribution, and was validated with simulated \0 events, resulting in a signal  efficiency of 99.7\%. This slightly larger-than-expected efficiency is attributed 
to the topology of \0 events, which are more localized than the $\gamma$ events used to produce the cut. This is the same reason why events in the figure appear slightly off-centered from the diagonal lines, closer to the top line. The tail at the 2615~keV  \isotope{Tl}{208} peak arises from the cascade of $\gamma$ rays in that specific decay and interactions in the skin LXe. These events are not expected to impact the \0 sensitivity because the tail is at high energies and does not contribute  at energies near $\Q$.

\begin{figure}[t]
    \centering
    \includegraphics[width=\columnwidth]{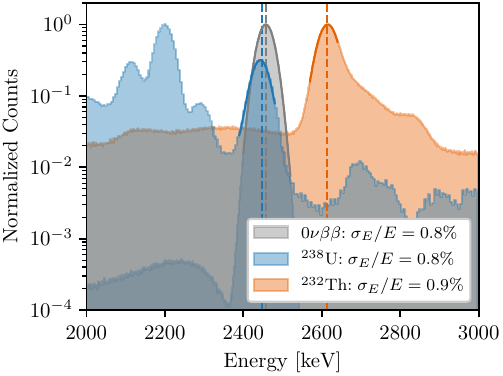}
    \caption{Rotated energy spectra of all simulated events in the FV from \0 decays and \isotope{Th}{232} and \isotope{U}{238} decays on the TPC vessel. The vertical lines indicate the expected center of the peaks while the solid lines show their Gaussian fits. The resolutions are calculated by the fitted $\sigma_E/E$. The three spectra are normalized to have the same maximum value.}
    \label{fig:energy-resol}
\end{figure}

The event energy is calculated by $E = (\tilde{N}_P + \tilde{N}_Q)\cdot W_q$, 
where $W_q$ is the average energy to create a single quantum of either charge or light. This approach is equivalent to a $-45^\circ$ projection of the event energies onto the rotated energy axis shown in Figure~\ref{fig:charge_light_cut}, which is nearly optimal for the expected noise in nEXO. Figure~\ref{fig:energy-resol} shows the rotated energy for simulated events from \isotope{Th}{232} and \isotope{U}{238} decays in the TPC vessel as well as from \0 decays in the LXe. The energy resolution extracted at the peak energies are $\sigma_E/E \simeq 0.8\%$. These values were found to be similar for $\gamma$ events with single or multiple energy deposits. These results are consistent with the model described in Ref.~\cite{nEXOpCDR}, and    
validated with EXO-200 data~\cite{PhysRevC.101.065501}. Studies of the dependence of the energy resolution on the electron lifetime show that a \SI{10}{\milli\second}, \SI{7}{\milli\second} or \SI{5}{\milli\second} lifetime calibrated to \SI{10}{\percent}, \SI{6}{\percent} or \SI{3}{\percent} accuracy, respectively, would reduce the overall energy resolution by only \SI{0.03}{\percent}, in absolute value, making its contribution subdominant. The effect of the energy resolution on the nEXO \0 sensitivity is discussed in section~\ref{sec:results}. 

\subsection{Deep neural network discrimination}

Background events in nEXO are expected to be dominated by $\gamma$-rays released from decaying nuclides present in and around the detector region. Their interactions in the LXe are primarily Compton scatterings which have the distinct feature of depositing energy in multiple locations, in contrast to $\beta$ or double-$\beta$ decays that predominantly leave localized deposits ranging only a few millimeters. 
On the other hand, at the energy of interest, bremsstrahlung radiation from fast electrons produced in \2 or \0 decays can make such events appear multi-sited and thus difficult to distinguish from gamma backgrounds. Motivated by the recent results of EXO-200~\cite{PhysRevLett.123.161802}, as well as analyses using nEXO simulations~\cite{Li_2019}, the simulated waveforms are used directly to develop a DNN that focuses on these topological differences in order to discriminate 
between double-$\beta$ and $\gamma$ events. The input to the network are two-dimensional time series for strips along the $x$ and $y$ directions. In this input, we also include simulated channels without charge collection but having induction signal height larger than 800~e$^-$. The algorithm implementation is developed with an 18-layer-deep residual network (ResNet-18)~\cite{7780459} using the PyTorch framework~\cite{paszke2017automatic}. 

The training data consists of two classes of 1.4~million simulated interactions each: \0 decay-like events with the sum energy of the two final-state electrons scaled to a uniform distribution between 900~keV and 3600~keV; and $\gamma$ rays with uniform energy in the same range. The location of the simulated decays and $\gamma$ interactions are drawn uniformly within the field cage. While \0 events occur within a very narrow energy range, the training of the DNN with a broad energy spectrum  minimizes the correlation between this variable and other background discriminators such as event energy and absolute location in the TPC. It also emphasizes the topological differences between double-$\beta$ decays and Compton scatterings that are independent of the background source. This dataset is randomly split, with 85\% of events used for training and 15\% for validation.  

\begin{figure}[t]
    \centering
    \includegraphics[width=\columnwidth]{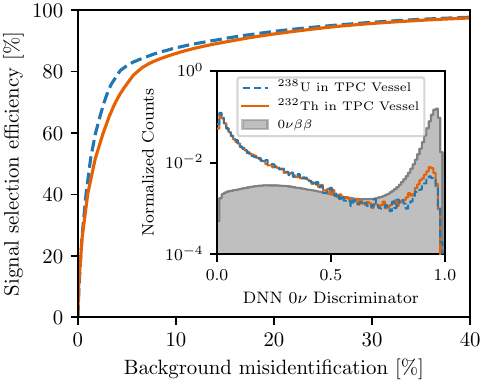}
    \caption{\0 decay signal selection efficiency versus background misidentification obtained with the DNN discriminator for simulated events whose reconstructed energy is within the FWHM around \Q. The solid orange (dashed blue) line refers to the \isotope{U}{238} (\isotope{Th}{232}) backgrounds from the TPC vessel. The inset shows the DNN \0 discriminator distribution for these events.}
    \label{fig:dnn_roc}
\end{figure}

Figure~\ref{fig:dnn_roc} shows the separation achieved between simulated events from \isotope{U}{238} and \isotope{Th}{232} $\gamma$-rays and \0 decays, with event energy within the full width at half maximum (FWHM) around $\Q$, after training and validation are complete. A \0 decay selection efficiency of $\sim80\%$ is achieved with only $\sim5\%$ misidentification of \isotope{U}{238} events. The rejection of \isotope{Th}{232} events is slightly worse because of the topology of its events overlapping with $\Q$. Almost 50\% of these events are in the outermost 1500~kg of the 3648~kg TPC LXe volume, and consist of energy deposited near the Compton edge for the \isotope{Tl}{208} 2615 keV $\gamma$ ray \cite{stiegler2020event}. As such, a larger fraction of these events consist of just a single recoiling electron rather than multiple energy deposits, and therefore more closely mimic \0 single-site interactions than events reconstructed in the full absorption peak of \isotope{Bi}{214} 2448 keV line (which can be made up of multiple short-range Compton scatterings).

The DNN performance was verified to be stable for reasonable ranges of the electronic noise in the waveforms. In particular, we observe that the obtained performance is equivalent to the separation used in Ref.~\cite{PhysRevC.97.065503} for a binary cut on the DNN discriminator of 0.85, but with an additional 17\% increase in signal detection efficiency. This result suggests a tagging procedure where scores above this value are classified as events with a single energy deposit, referred to as ``single-site'' (SS-like), and those below the cut contain multiple deposits, \emph{i.e.}, ``multi-site'' (MS-like) events.

\section{Background Model}\label{sec:materials}

A credible background estimation, founded on radioassay data and coupled with a detailed MC model, is key to an accurate determination of the experiment performance. For this purpose, a comprehensive list of background sources must be built. This process was successfully demonstrated in EXO-200 where the predicted background was found to agree, within uncertainties, with the one derived from the final fit to the data \cite{PhysRevC.92.015503}. The results presented here are based on a robust bottom-up estimation of the backgrounds from individual components and we note that, as far as intrinsic radioactivity is concerned, these results do not involve any extrapolation of materials radiopurity beyond what has been already measured.

Background sources are classified into three main categories for nEXO: intrinsic radioactivity of detector materials, radon outgassing, and radionuclides that form or are introduced during the course of the material production, detector assembly, and operation. Only background components resulting in more than 0.01 SS-like events/(FWHM$\cdot$yr) in the central 2000~kg are considered here. The background rate in this central mass region provides a simplified reference value for the experiment's performance. This is because an analysis performed with events in that region yields $\sim 95\%$ of the experimental sensitivity achieved with the full FV, as presented in Ref.~\cite{PhysRevC.97.065503} and re-validated for this work, although the full mass is required for self-shielding and measuring the background rates.  

\subsection{Intrinsic radioactivity of detector materials}

\begin{table*}[t!]
\centering
\small
\setlength{\tabcolsep}{6pt}
\renewcommand{\arraystretch}{1.15}%
\begin{tabular}{lllccccc}
\hline
\textbf{Material}     & \textbf{Supplier}               & \textbf{Method}  & \textbf{\isotope{U}{238}}     & \textbf{\isotope{Th}{232}}         & \textbf{\isotope{K}{40}}  & \textbf{\isotope{Co}{60}}     \\ 
     &               &   & \textbf{ (ppt)  }     & \textbf{ (ppt)}         & \textbf{(ppb)}  & \textbf{(mBq/kg)}     \\ 
\hline
Electroformed Copper$^\dagger$ & In-house & ICP-MS & $< 0.01$ & $< 0.01$ & - & - \\ 
Copper & Aurubis & ICP-MS/Ge/ & $0.254 \pm 0.008$ & $0.13 \pm 0.06$ & $< 6.4$ & $< 0.0033$ \\
 &  & GDMS &  &  &  \\
Sapphire & GTAT & NAA & $< 8.9$ & $6.0 \pm 1.1$ & $9.5 \pm 2.0$ & - \\
Quartz & Heraeus & NAA & $< 1.5$ & $< 0.23$ & $0.55 \pm 0.03$ & - \\
SiPM & FBK & ICP-MS/NAA & $0.86 \pm 0.05$ & $0.45 \pm 0.12$ & - & - \\
Epoxy$^*\text{}^\dagger$ & MasterBond & Ge & $< 360$ & $< 540$ & $< 930$ & - \\
AuSn solder$^\dagger$ & Nippon Micrometal & ICP-MS & $90 \pm 20$ & $68 \pm 14$ & - & - \\
Gold wire bonding$^\dagger$ & Ametek & ICP-MS & $< 230$ & $26 \pm 8$ & - & - \\
Polyimide$^\dagger$ & Taiflex & ICP-MS & $0.71 \pm 0.04$  & $0.71 \pm 0.20$ & - & - \\
 &  &  & pg/cm$^2$ & pg/cm$^2$ &  &  \\
HFE-7000$^*$ & 3M & NAA & $< 0.015$ & $< 0.015$ & - & - \\ 
CFC (Resin)$^\dagger$ & SCC & Ge  & $< 7.7$ & $< 19$ & $< 31$ & $< 0.03$ \\
CFC (Fiber)$^\dagger$ & Grafil & Ge  & $40 \pm 15$ & $74 \pm 39$ & $810 \pm 100$ & $< 0.11$ \\
ASICs (Silicon)$^\dagger$ & Global Foundry  & ICP-MS & $0.35 \pm 0.13$ & $1.3 \pm 0.7$ & - & - \\ 
Titanium$^\dagger$ & LZ TIMET & Ge  & $ <12$ & $57 \pm 5$ & $ <29$ & $ <0.033$ \\
Water & SNOLAB &  & $< 1$ & $< 1$ & $<1000$ & - \\
Stainless Steel$^\dagger$ & GERDA & Ge  & $ <48$ & $ <200$ & $ <58$ & $16.8 \pm 2.4$ \\
HDPE$^\dagger$ & Dielectric Sci. & NAA & $100 \pm 19$ & $63.6 \pm 2.7$ & $350 \pm 10$ & - \\
PTFE$^*\text{}^\dagger$ & DuPont & NAA & $ <0.78$ & $ <0.26$ & $1.8 \pm 0.2$ & - \\
Cond. PE$^\dagger$ & Quadrant & ICP-MS & $224 \pm 32$ & $10.1 \pm 1.4$ & - & - \\
\hline
\end{tabular}
\caption{Materials, analysis method and radioactivity concentrations entering the nEXO background model. Data for entries marked with a $^*$ were taken from the EXO-200 materials certification program. Entries marked with a $^\dagger$ are either new or updated from the previous analysis~\cite{PhysRevC.97.065503}. Data for titanium, water and stainless steel are from Table~VI of Ref.~\cite{Akerib:2017iwt}, Ref.~\cite{Aharmim:2011yq} and from Ref.~\cite{Maneschg:2008}, respectively. Limits are stated at 90\% CL and were computed using the ``flip-flop'' method \cite{Feldman:1997qc}.}
\label{tab:rad}
\end{table*}

The long-lived radionuclides of primary interest for nEXO are \isotope{U}{238}, with a $\gamma$-ray at 2448~keV (from   \isotope{Bi}{214} decay) and \isotope{Th}{232}, with a $\gamma$-ray at 2615~keV (from \isotope{Tl}{208} decay). 
Secular equilibrium is assumed for the decay chains. While \isotope{Co}{60} and \isotope{K}{40} are kept in the background model because of their impact on the measurement of \2 decays, \isotope{Cs}{137} is excluded since interactions depositing $<700$~keV in the TPC are discarded in the analysis. \isotope{Al}{26}, with its $\gamma$-ray emission at 2938~keV, is included in the model since it could be present in the sapphire part of the FSR support at levels of $<0.9$~mBq/kg at 90\% CL. The mass of Al contained in the reflective coatings is very small, so that its contribution to the \isotope{Al}{26} activity is negligible. 

A list of radioassay-based input activity levels for nEXO materials is provided in Table~\ref{tab:rad}. The radioassay effort in nEXO is  conducted at multiple collaborating labs worldwide, with cross-calibration efforts to ensure standardization and uniformity in the results. The various techniques utilized are detailed in Ref.~\cite{nEXOpCDR,PhysRevC.97.065503}. In addition to the new entries due to the refined implementation of the geometry in MC simulations, radioactivity measurements for polyimide in the readout cables and for the ASICs in the electronics were updated from Ref.~\cite{PhysRevC.97.065503}. 
Previously unpublished radioactivity data for the CFC has been utilized in the background estimate for the OV and IV. Electrical connections in the TPC and to the cables bringing signals out of it use gold, epoxy and gold-tin alloy, reproducing the current understanding of the design. Following substantial engineering feasibility studies and  R\&D on in-house copper electroforming~\cite{Hoppe:2014nva,Abgrall:2016cct}, it was concluded that direct electroforming the TPC cylindrical vessel and other components is possible. 
Therefore this technology of electroformed (EF) copper is now the baseline for nEXO and results in a significantly lower background.

\begin{table*}[tbp]
\centering
\setlength{\tabcolsep}{6pt}
\renewcommand{\arraystretch}{1.15}%
\begin{tabular}{lllc}
\hline
\textbf{Component}    & \textbf{Nuclides}                                      & \textbf{Material} & \textbf{Mass or}              \\ 
& \textbf{Simulated} & & \textbf{Surface Area}\\
\hline 
Outer Vessel Tank Support$^\dagger$ & \isotope{U}{238}, \isotope{Th}{232} & Stainless Steel  &  2711 kg                                \\
Outer Vessel Support$^\dagger$ & \isotope{U}{238}, \isotope{Th}{232}, \isotope{K}{40}, \isotope{Co}{60} & CFC  &  690 kg                                \\
Outer Vessel        & \isotope{U}{238}, \isotope{Th}{232}, \isotope{K}{40}, \isotope{Co}{60} & CFC  &  2700 kg                                \\
Outer Vessel  Feedthroughs$^\dagger$      & \isotope{U}{238}, \isotope{Th}{232}, \isotope{K}{40}, \isotope{Co}{60} & Stainless Steel  & 12.5  kg                                \\
Outer Vessel Liner$^\dagger$   & \isotope{U}{238}, \isotope{Th}{232}, \isotope{Co}{60}              & Titanium          &  858.8 kg                             \\
Inner Vessel Support$^\dagger$ & \isotope{U}{238}, \isotope{Th}{232}, \isotope{K}{40}, \isotope{Co}{60} & CFC  &  447 kg                                \\
Inner Vessel        & \isotope{U}{238}, \isotope{Th}{232}, \isotope{K}{40}, \isotope{Co}{60} & CFC  &  823 kg                                 \\
Inner Vessel Liner   & \isotope{U}{238}, \isotope{Th}{232}, \isotope{K}{40}, \isotope{Co}{60}             & Titanium          &  323.4 kg                             \\
Inner Vessel Feedthroughs$^\dagger$   & \isotope{U}{238}, \isotope{Th}{232}, \isotope{K}{40}, \isotope{Co}{60}             & Titanium          &  8 kg                             \\
HFE-7000                   & \isotope{U}{238}, \isotope{Th}{232}, \isotope{Rn}{222}   & HFE-7000          & 31814 kg      \\
TPC Vessel Support$^\dagger$     & \isotope{U}{238}, \isotope{Th}{232}, \isotope{K}{40}, \isotope{Co}{60}                            & Copper            & 83 kg             \\
TPC Vessel            & \isotope{U}{238}, \isotope{Th}{232}, \isotope{K}{40}, \isotope{Co}{60}                            & Copper            & 447 kg                              \\
FSRs     & \isotope{U}{238}, \isotope{Th}{232}, \isotope{K}{40}, \isotope{Co}{60}                            & Copper            & 68 kg                               \\
FSR Support Components   & \isotope{U}{238}, \isotope{Th}{232}, \isotope{K}{40}, \isotope{Al}{26}              & Sapphire          & 2.59 kg                                      \\
Cathode              & \isotope{U}{238}, \isotope{Th}{232}, \isotope{K}{40}, \isotope{Co}{60}                            & Copper            & 12.18 kg                               \\
SiPM                  & \isotope{U}{238}, \isotope{Th}{232}               & SiPM              & 2.91 kg                               \\
SiPM Staves           & \isotope{U}{238}, \isotope{Th}{232}, \isotope{K}{40}, \isotope{Co}{60}                            & Copper            & 132.4 kg                              \\
SiPM Module Backing   & \isotope{U}{238}, \isotope{Th}{232}, \isotope{K}{40}                            & Quartz            & 11.23 kg                                      \\
SiPM Electronics      & \isotope{U}{238}, \isotope{Th}{232}                            & ASICs           & 2.2 kg                                      \\
SiPM Module Wire bonds       & \isotope{U}{238}, \isotope{Th}{232}               & Gold          & 8 g                               \\
SiPM Cables           & \isotope{U}{238}, \isotope{Th}{232}                            & Polyimide            & $1.53\times 10^4$ cm$^2$ \\
SiPM Bump Bonds$^\dagger$  & \isotope{U}{238}, \isotope{Th}{232}                            & AuSn            & 2.2 g \\
Charge Module Support & \isotope{U}{238}, \isotope{Th}{232}, \isotope{K}{40}, \isotope{Co}{60} & Copper            & 34.12 kg                               \\
Charge Module Backing & \isotope{U}{238}, \isotope{Th}{232}, \isotope{K}{40}  & Quartz            & 1.22 kg                               \\
Charge Module Electronics    & \isotope{U}{238}, \isotope{Th}{232}                            & ASICs           & 0.42 kg                    \\
Charge Module Cables  & \isotope{U}{238}, \isotope{Th}{232}                            & Polyimide            & $0.63\times 10^4$ cm$^2$                \\
Charge Module Wire bonds       & \isotope{U}{238}, \isotope{Th}{232}              & Gold          & 2 g                               \\
Charge Module Epoxy  & \isotope{U}{238}, \isotope{Th}{232}                            & Epoxy            & 1 g \\
HV Feedthrough Components$^\dagger$   & \isotope{U}{238}, \isotope{Th}{232}, \isotope{K}{40}, \isotope{Co}{60}                            & Copper            & 7.95 kg \\
HV Feedthrough Components$^\dagger$   & \isotope{U}{238}, \isotope{Th}{232}, \isotope{K}{40}                            & PTFE              & 0.71 kg \\
HV Cable$^\dagger$               & \isotope{U}{238}, \isotope{Th}{232}, \isotope{K}{40}                            & Conductive PE            & 2 g \\
HV Cable$^\dagger$               & \isotope{U}{238}, \isotope{Th}{232}, \isotope{K}{40}                            & HDPE            & 0.55 kg \\
Calibration Guide Tubes$^\dagger$      & \isotope{U}{238}, \isotope{Th}{232}, \isotope{K}{40}, \isotope{Co}{60}           & Copper            & 4.92 kg \\
TPC LXe Volume     & \isotope{Xe}{137}, \isotope{Rn}{222},\isotope{Ar}{42}, \2, \0                   & Xenon             & 3648 kg                   \\
Skin LXe Volume                   & \isotope{Xe}{137}, \isotope{Rn}{222},\isotope{Ar}{42}, \2, \0                   & Xenon             & 1163 kg                  \\ 
\hline
\end{tabular}
\caption{List of simulated detector components of the nEXO background model that is used in the sensitivity calculations, along with their material, nuclides simulated, and mass or surface area. Relative to Ref.~\cite{PhysRevC.97.065503}, components marked with $^\dagger$ were introduced in this analysis, as well as simulations of \isotope{Co}{60} decays, \isotope{Rn}{222} in the HFE-7000, and \isotope{Ar}{42} in the LXe volume. Components simulated but found to contribute negligibly to the background are not listed.}
\label{tab:components}
\end{table*}

Table~\ref{tab:components} shows a complete list of the simulated detector components used in the background model of the sensitivity analysis. Pure $\beta$ emitters on TPC surfaces are not included since the short range of $\beta$ particles in LXe prevents them from passing the FV selection cut. 


Isotopic enrichment efficiently suppresses all species lighter than 136 amu with exception of 134 amu, which makes up the remaining 10\% LXe mass. \isotope{Xe}{134} double-$\beta$ decays with $Q$-value of $825.8\pm0.9$~keV~\cite{Wang2012} and is not a background for \isotope{Xe}{136}. Continuous xenon recirculation and purification guarantees the extreme purity of the LXe in the detector. While only the direct measurement with the nEXO detector will be able to confirm that the background from impurities in the xenon is negligible, we consider for the purpose of this estimate only  intrinsic contaminations that can plausibly be problematic.  These are the \isotope{Xe}{136}, which undergoes $2\nu\beta\beta$,  and \isotope{Ar}{42}. The \2 decay rate is based on the EXO-200 half-life measurement of $(2.165\pm0.061) \times 10^{21}$~yr~\cite{PhysRevC.89.015502}. The very long half life and the energy resolution of nEXO make this background negligible, even at the largest exposure considered here.

Argon is present in xenon as both are obtained as by-product of atmospheric gas distillation. Among its  long-lived isotopes, only \isotope{Ar}{42} is a concern for nEXO. The decay of its daughter \isotope{K}{42} results in a $\beta$ emission with endpoint energy of 3525~keV, as well as $\gamma$ radiation with an energy of 2424~keV (with a branching fraction of 0.02\%). Multiple factors play a role in estimating the expected levels of \isotope{Ar}{42} in nEXO, some of which are not well known, including the  abundance of  \isotope{Ar}{42} relative to \isotope{Ar}{40} after enrichment, which is difficult to estimate quantitatively because they are both very far from the mass cut used. Here we assume a factor nine enhancement of the isotope \isotope{Ar}{42} abundance based on the equivalent value for \isotope{Xe}{136} separation relative to lighter xenon isotopes. Based on measurements of \isotope{Ar}{40} in enriched xenon by the EXO-200 collaboration \cite{Dobi:2011zx}, the assumptions of a xenon recirculation time of 4 days and 100\% efficient removal of \isotope{K}{42} by the purifier, and an ionization fraction of 76\% for this isotope (based on measurements of Rn daughters in LXe~\cite{Albert:2015vma}), we obtain an internal \isotope{K}{42} activity of $2.59\times10^{-9}$~mBq/kg, which we include in the model. Even assuming a five times larger activity, the impact on the sensitivity is  negligible. 

MC studies confirmed that the background contribution from natural radioactivity in SNOLAB's cryopit rock walls, including those from the layers of concrete and shotcrete result in negligible background contribution and have, therefore, been omitted from the model. Similarly, the contributions from \isotope{U}{238} and \isotope{Th}{232} present in the water and the PMTs of the muon veto system can be neglected even if these nuclides leak into the water and are present near the OV surface, which is a conservative assumption.

\subsection{Radon outgassing}
\label{sec:radon}

As the background from the copper is substantially reduced compared to previous estimates, contributing to the better projected sensitivity, \isotope{Rn}{222} in the xenon contributes a larger share and is modeled in detail.  The current estimate of the  \isotope{Rn}{222} background is experimentally constrained by EXO-200 data, 
taking into account the emanation from the various detector and recirculation-system components and their relative surface areas. This process results in a steady-state population of $N_\text{Rn} = 600$ \isotope{Rn}{222} atoms contained in the xenon. 

A number of options exist to mitigate the radon content of the xenon, including the selection of materials and components based on their \isotope{Rn}{222} emanation rate and the development of surface cleaning protocols. This must be done for all components wetting the xenon. EXO-200 data showed steady-state population of about 200 \isotope{Rn}{222} atoms present in the TPC throughout the physics runs. These data also indicated that the likely source of this radon was external to the TPC. 
To identify the principal source, the \isotope{Rn}{222} production rate of the EXO-200 xenon handling system (piping, pump, purifier) is being radon-assayed in various configurations. A radon background budget will be kept as components are accepted for installation in the detector, and transport models will be used to maintain current predictions of the radon population in the TPC. The xenon handling system will be tested for its \isotope{Rn}{222} emanation rate during construction to confirm that its components do not exceed the requirements. 
Ongoing measurements appear to indicate that much of the emanation in EXO-200 occurred in the xenon purifiers. Work is in progress to better understand and reduce such emanation. 
One of the strengths of a liquid-phase detector such as nEXO is that the external recirculation and purification system can be upgraded, should the need arise. A distillation system could be utilized to reduce this contamination to acceptable levels. The effectiveness of this approach has been demonstrated in previous work \cite{Aprile:2017kop,Cui:2020bwf}.

\begin{table}[tbp]
\centering
\setlength{\tabcolsep}{7pt}
\renewcommand{\arraystretch}{1.15}%
\begin{tabular}{l|cc|c}
\hline
Decay Location $\ell$ & $f_\ell$  & $\varepsilon_{\alpha,\ell}$ & $N_{\text{Rn},\ell}$ \\
\hline
Cathode top surface & 0.617& 0.5 & 185.41 \\
Field rings surfaces & 0.184 & 0.49 & 56.32 \\
Cathode bottom surface & 0.044 & 0.01 & 26.37 \\
Skin LXe above anode & 0.012 & 0.01 & 7.04 \\
Skin LXe under cathode & 0.003 & 0.01 & 1.65 \\
TPC LXe bulk & 0.128 & 0.999 & 0.21 \\
Skin LXe barrel & 0.012 & 0.98 & 0.16 \\
\hline
\end{tabular}
\caption{List of \isotope{Bi}{214} decay locations $\ell$ with the corresponding decay fractions $f_\ell$, $\alpha$ tagging efficiency $\varepsilon_{\alpha,\ell}$, and number of background-contributing atoms $N_{\text{Rn},\ell}$ present in steady-state in the LXe.}
\label{tab:radon}
\end{table}

In the simulations and data analysis, the treatment of \isotope{Rn}{222} in the LXe follows the methodology described in Ref.~\cite{stiegler2020event}.
$\beta$ emission from \isotope{Bi}{214} is followed in fast succession by a \isotope{Po}{214} $\alpha$ decay that can be used to identify the Bi-Po decay chain if the decays are fully contained in the LXe. The efficiency of this approach depends on the choice for the coincidence time window and the location of the $\alpha$ decay. Table~\ref{tab:radon} shows the fraction ($f_\ell$) of radon-related \isotope{Bi}{214} decays  in seven detector areas $\ell$ along with the  $\alpha$ tagging efficiency  $\varepsilon_{\alpha,\ell}$~\cite{stiegler2020event}. The \isotope{Bi}{214} decays have to be partitioned this way to account for drift of ionized \isotope{Rn}{222} daughters in different electric field regions of the LXe volume. For a veto time of $\Delta t_\text{BiPo}=1500$~$\mu$s and a \isotope{Po}{214} half life of 164~$\mu$s, 
the Bi-Po coincidence tagging efficiency is $\varepsilon_\text{BiPo} =99.82\%$. 
The number of untagged \isotope{Bi}{214} decays at each location is  $N_{\text{Rn},\ell} = N_\text{Rn} \cdot f_\ell \cdot (1-\varepsilon_\text{BiPo}\cdot\varepsilon_{\alpha,\ell})$. 

After application of the BiPo veto,  approximately 46\% (277 atoms) of the steady-state population of \isotope{Rn}{222} atoms in the LXe contribute to the background. Most of the \isotope{Bi}{214} decays from the \isotope{Rn}{222} chain are from regions at the edges of the TPC, allowing for optimal exploitation of the standoff and single-site discriminators, and keeping the innermost LXe volume radioquiet while allowing measurement of this background in the outer volume.

\isotope{Rn}{222} in the HFE-7000 will also result in \isotope{Bi}{214} decay-related backgrounds, with the relevant emanation primarily originating from the titanium liner of the IV. Using measurements from Ref.~\cite{RadonLZ,RadonXenon1T}, this is expected to result in a steady-state population of $\sim1500$ atoms of  \isotope{Rn}{222} in the HFE-7000.  The baseline design of the cryogenic system assumes recirculation of the HFE-7000 in an external heat exchanger. It is assumed that the contribution from such an external system can be kept subdominant to the titanium liner contribution due to its much smaller surface area. The current simulation assumes 1500 Rn atoms uniformly distributed throughout the HFE-7000 volume. The same \isotope{Rn}{222} mitigation strategies described for the LXe system apply to HFE-7000 system. While this source was added to the model, \isotope{Rn}{222} in the OD water was not included because it can be readily maintained below $9\times10^{-9}$~Bq/kg~\cite{Aharmim:2011yq} and therefore is a negligible contribution to the background budget. 

\subsection{Exposure-based backgrounds}
\label{sec:exposure-based-bkg}

The backgrounds considered so far are primarily related to contaminants in the material bulk and intrinsic radon outgassing level. 
Background also arises from the exposure of detector components to cosmogenic radiation, to radon in the air, and to dust. These background contributions are thus a function of the material handling and detector operation.  

For cosmogenically-created radionuclides, two separate production pathways are considered: activation while materials are fabricated, transported, or stored above ground and steady-state production underground. 

For above-ground activation, a systematic study of possible radionuclides produced by cosmogenic activity, including production rates and hit efficiencies, is currently underway for all detector materials. For the initial effort, the background contribution from cosmogenic isotopes in copper (e.g. \isotope{Co}{56}, \isotope{Co}{60}, \isotope{Sc}{44}, \isotope{K}{42}) was evaluated and found to be acceptable with proper management of above-ground exposure and a sufficient period of cool-down underground before deployment. In addition, xenon spallation products that can be removed by purification are not a concern for nEXO.

Underground cosmogenic activation of detector components was evaluated in a prior work \cite{PhysRevC.97.065503} using MC simulations. The resulting backgrounds were found to be negligible with the notable exception of those arising from \isotope{Xe}{137}, with a half-life of  $T_{1/2}^{^{137}\text{Xe}}=3.82$~min~\cite{CARLSON1969267}. In particular, \isotope{Xe}{135} produced through neutron capture in \isotope{Xe}{134} has an endpoint energy of $1165$~keV~\cite{SINGH2008517} and is not a background for \isotope{Xe}{136}. The $\beta$ emission from \isotope{Xe}{137} decay has an endpoint energy of $4173$~keV~\cite{Browne20072173} and this causes an overlap of the $\beta$ spectrum in the region of the $\Q$ value. 
The contribution from \isotope{Xe}{137}, activated by interactions in the bulk of LXe, was studied using \geant{} MC simulations with $\sim 10^{7}$~muons passing through a cylinder coaxial with the water tank, but with slightly larger dimensions than the OD size, and following energy and angular distributions observed at SNOLAB~\cite{PhysRevD.73.053004}. Secondary particles were tracked and the number of \isotope{Xe}{137} atoms produced from neutron captures was counted.  Nuclei are produced with an excited state energy of  $4026$~keV~\cite{NSR2012WA38}, which is promptly released through $\gamma$ emission. These prompt emissions provide a tag for \isotope{Xe}{137} production by coincidence with the originating muon. The veto efficiency was estimated from the muon tagging efficiency and from the fraction of \isotope{Xe}{136} neutron-capture events reconstructed with energy within $\pm2\cdot$FWHM of the full de-excitation energy. A 70\% veto efficiency was obtained for \isotope{Xe}{137} decays near the FV edge, increasing towards the LXe center. Therefore, this value is conservatively applied to reduce the background contribution from \isotope{Xe}{137} to $(0.85\pm 0.04)\times 10^{-3}$~atoms/(kg$\cdot$yr), uniformly distributed in the LXe volume,  without significantly decreasing the livetime, even for a veto time window as long as five \isotope{Xe}{137} half-lives.

Production of \isotope{Xe}{137} could also arise from capture of neutrons not of cosmogenic origin. Radiogenic neutrons produced in the cavern walls by $(\alpha,n)$ reactions in the rock typically have energies lower than 10~MeV, and do not reach the TPC due to attenuation by the water shielding. However, activation of \isotope{Xe}{136} during xenon recirculation outside the water tank was evaluated and found to be non-negligible unless special precautions are taken. MC studies indicate that \isotope{Xe}{137} production in these regions can be mitigated by ensuring sufficient neutron shielding (e.g. a few cm of borated polyethlyene) of the circulation pipes outside the OD. Such shielding is being incorporated into the design of the recirculation system. Fast neutrons can also be created close to the TPC via $(\alpha, n)$ reactions by $\alpha$-decays from the natural decay series and out-of-equilibrium $^{210}$Po (e.g., from \isotope{Rn}{222} daughter plate out) from components inside the water shield. Spontaneous fission contributes relatively little to this background. Cross sections for ($\alpha$, n) reactions at these energies are relevant only in low-$Z$ materials due to the nuclear Coulomb barrier, meaning only certain materials are of particular concern, as well as \isotope{Rn}{222} exposure during detector construction. MC studies indicate that when U and Th decay chains are the source for these $\alpha$ particles, the resulting \isotope{Xe}{137} production was found to be 100 to 1000 times smaller than the background contributions from $\gamma$-rays of the same chain. Under the assumption that the introduction of these backgrounds can be sufficiently controlled during detector design and construction, these processes are neglected in the current model. Radon progeny attachment studies are underway to provide the underpinning for formulating radon reduction specifications for the detector assembly areas. Radon removal devices will be employed during construction to meet this requirement.

Dust can contribute to the nEXO background in two different ways: emission of $\gamma$-radiation and, in case it is in contact with the xenon, emanation of \isotope{Rn}{222}. The fractional contribution of either component depends on the, yet unknown, fraction of mobile dust particles. Another source of uncertainty is the make-up of the dust particles. The collaboration developed a dust fallout model to guide this discussion. During construction dust will be measured via witness plates and tape lifts ~\cite{diVacri:2020aqc}. Particles can then be counted via optical microscopy and their U/Th contents can be measured by inductively coupled plasma mass spectrometry (ICP-MS). nEXO plans to organize its work around improved EXO-200 assembly procedures. The low \isotope{Rn}{222} content of the xenon in EXO-200 is indirect evidence for the success of its cleanliness protocols. In the current sensitivity estimate it is assumed that dust mainly contributes via \isotope{Rn}{222} emanation. Its contribution is absorbed in the overall \isotope{Rn}{222} content presented in section~\ref{sec:radon}.

During detector operation, backgrounds from interactions of neutrinos must also be considered. Among these, the only important contribution arises from electron-neutrino elastic scattering from \isotope{B}{8} solar neutrinos, uniformly interacting throughout the LXe volume~\cite{PhysRevC.97.065503}. This scattering results in $(1.15\pm 0.06)\times 10^{-3}$~electrons/(kg$\cdot$yr) with a continuous energy spectrum having an end point that is above $\Q$~\cite{de_Barros_2011}. This contribution is included in the MC model.

\subsection{Background budget}

\begin{figure}[tbp]
\centering
\includegraphics[width=0.99\columnwidth]{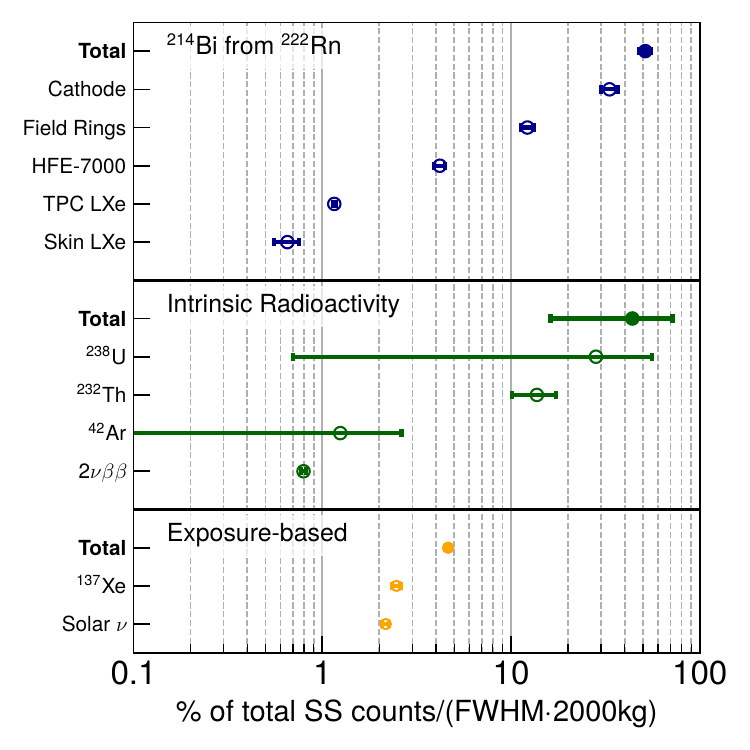}
\caption{SS-like fractional background contributions with energy within \Q$\pm$FWHM/2 and in the inner 2000 kg. The contributions are grouped by category, as described in the text. For each category, the total contribution is shown by the solid marker, while individual contributions are indicated by open circles. Negligible contributions are not shown. For \isotope{Rn}{222} backgrounds, the breakdown by \isotope{Bi}{214} decay location (based on Table~\ref{tab:radon}) is shown. Breakdown by the individual source terms is given for the other two background categories.}
\label{fig:nEXO_background_budget_overview}
\end{figure}

\begin{figure*}[tbp]
\centering
\includegraphics[width=0.49\textwidth]{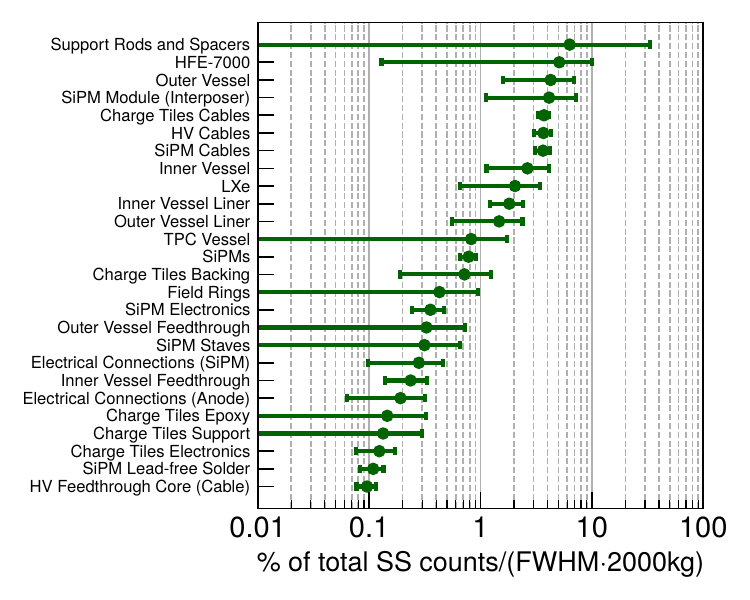}
\includegraphics[width=0.49\textwidth]{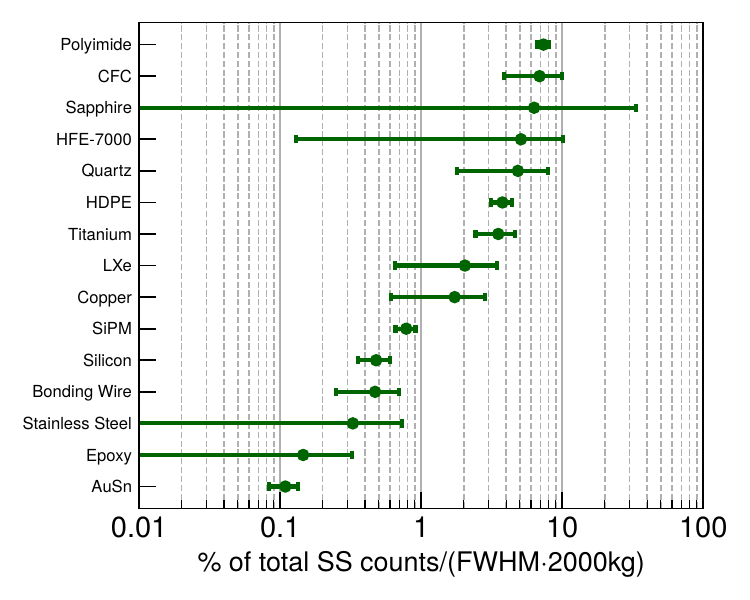}
\caption{Fractional background contributions from intrinsic radioactivity of the detector materials grouped by detector component (left) and material (right) with SS-like topology and energy within \Q$\pm$FWHM/2 and in the inner 2000 kg.}
\label{fig:nEXO_background_budgets}
\end{figure*}

The development of a background budget is valuable for understanding the role of different contributions, optimizing the design, planning contingencies, and focusing resources during detector construction. In order to capture the complex impact of the different background contributions in a synthetic manner, a representative value is computed for each component and their relative contributions are studied. MC simulations are used to evaluate the probability that a decay in a specific detector component will produce an SS-like event, \emph{i.e.}, an event with \0 DNN discriminator $>0.85$, with all energy deposits reconstructed within the inner 2000~kg LXe volume, and event energy within the FWHM of $\Q$~\cite{PhysRevC.97.065503}. These probabilities, referred to as the hit efficiencies, are then combined with the radioactivity content from each material 
using a truncated Gaussian distribution to model the background contributions~\cite{tsang2018treatment}. Uncertainties are propagated from the underlying terms. An overview of the relative contributions from each background category is shown in Figure~\ref{fig:nEXO_background_budget_overview}. For radon outgassing background, only the statistical uncertainty is given.
The breakdowns by material and components of the intrinsic background from radioactivity of nEXO's detector materials are shown in Figure~\ref{fig:nEXO_background_budgets}. 

In the current model, the largest background contribution is from \isotope{Rn}{222}, with $\sim50\%$ of the total budget, and more specifically from \isotope{Rn}{222} deposited on the cathode, which corresponds to $\sim30\%$ of the total budget. This result contrasts with previous nEXO publications where the copper for the TPC construction was dominant. This is the result of the decision to use in-house EF copper which has lower intrinsic activity. Indeed, in a scenario with all copper from the previously assumed commercial material, 
this material alone would consume $\sim50$\% of the total background budget, with the TPC vessel remaining the largest individual contributor with $\sim20\%$ of the total budget, instead of its current $<1\%$ contribution. 

Other radio-impurities contained in the current model of nEXO's detector materials account for about 45\% of the background, with exposure-based backgrounds comprising only $\sim5\%$ of the total. Among the backgrounds from intrinsic material contamination, no single component or material appears to dominate. The contribution from the top four components comes with significant uncertainties from the underlying radioassay measurements, with the sapphire entry being particularly noticeable. Efforts are ongoing to improve these assays results, in many cases pushing the limits of available experimental sensitivity. More broadly, the background analysis shown here has been guiding the detector design by informing and prioritizing component design, material selection, radioassay, and overall detector optimization. We discuss the effect from \isotope{Rn}{222} contamination and cosmogenic \isotope{Xe}{137} activation in the LXe volume on the nEXO \0 sensitivity  in section~\ref{sec:results}.

\section{Sensitivity and Discovery Potential to \0 Decay}\label{sec:sensitivity}

The experimental sensitivity to the \0 decay half-life of \isotope{Xe}{136} provides the fundamental metric to determine the experiment's physics reach and guide its design by embedding all the detailed aspects of the design, material selection, and analysis into a single value of interest. The discovery potential and half-life sensitivity of the nEXO experiment to \0 is evaluated in a frequentist approach using a binned profile likelihood ratio test. As such, it  mimics the analysis that would be performed on real data and maximally exploits the multiparameter measurement capabilities of nEXO.

The procedure to compute the sensitivity follows the methodology outlined in Ref.~\cite{PhysRevC.97.065503}. First, probability density functions for each component and background source are constructed from the MC simulations. These are weighted by the radioassay measurements to produce a model of the nEXO measurement. Toy datasets are then generated by sampling events from this model, which is fitted back to the toy data in order to generate confidence intervals on the signal strength. The resulting ensemble of confidence intervals under different signal hypotheses is used to determine either the median $3\sigma$ discovery potential or sensitivity at the 90\% CL to \0 decay with nEXO.

\subsection{Methodology}

We begin by applying selection criteria which mimic an analysis of real data. We select events with reconstructed energies between 1000 and 3500~keV, and require the reconstructed standoff distance to be consistent with the FV defined in Section~\ref{sec:simulation} (larger than $2$~cm). The ``diagonal cut,'' illustrated by the dashed lines in Figure~\ref{fig:charge_light_cut}, is then applied to remove events with excess light in the skin. In EXO-200 data, this cut was also  sufficient to remove any $\alpha$ decays as well as poorly reconstructed $\beta$ and $\gamma$ events. The detection efficiency for $0\nu\beta\beta$ decay events after applying these cuts is 96.0\%. 

Events passing the cuts are binned into histograms with three dimensions: event energy, standoff distance, and the DNN \0 discriminator. A nonuniform binning scheme was developed to employ finer granularity where the signal and background distributions are most different, in order to maximize the discrimination power. The histograms are normalized to produce a set of probability density functions (PDFs), each associated with a component and nuclide pair, that is used to model the experimental data and its analysis.

The relative weight of each PDF is given by:
\begin{equation}
    N_{\text{PDF}} = M \,A_s\, \left(\frac{K}{N_\text{prim}}\right) \, T_\text{live}
    \label{eq:expected_counts}
\end{equation}
where $M$ is the mass (or surface area) of the particular component, $A_s$ is the specific activity per unit mass (or surface area), $K$ are the number of events passing our analysis cuts, $N_\text{prim}$ is the number of primary decays generated in the simulation, and $T_\text{live}$ is the assumed livetime. The hit efficiencies are given by $K/N_\text{prim}$. By employing input parameters that derive from a detailed engineering design, a robust radioassay program, and detailed high-statistics MC simulations, we feel confident that the PDFs and their normalization provide an accurate representation of the final experiment.

In order to account for the uncertainties in the radioassay measurements, $A_s$ is sampled from a normal distribution truncated at zero each time a new toy experiment is run. Fluctuations are applied at the material level, meaning that two components made of the same material fluctuate together in each toy experiment. 

All PDFs for either \isotope{U}{238} or \isotope{Th}{232} decays in detector components inside the TPC vessel are grouped together because of the degeneracy in their distributions. The PDFs for \isotope{K}{40} of all components are grouped together to simplify the calculations, since this nuclide does not contribute with events near \Q. In addition, all PDFs at a distance farther than the TPC vessel are grouped into a single ``far component''. Backgrounds in the liquid xenon itself, specifically \isotope{Xe}{137} and \isotope{K}{42} decays, are grouped into a separate component. On the other hand, \2 and \isotope{Rn}{222} decays in the LXe are treated as individual components, as well as the PDF for electron-neutrino elastic scattering from \isotope{B}{8} solar neutrinos.

\begin{figure*}[tp]
    \centering
    \includegraphics[width=\textwidth]{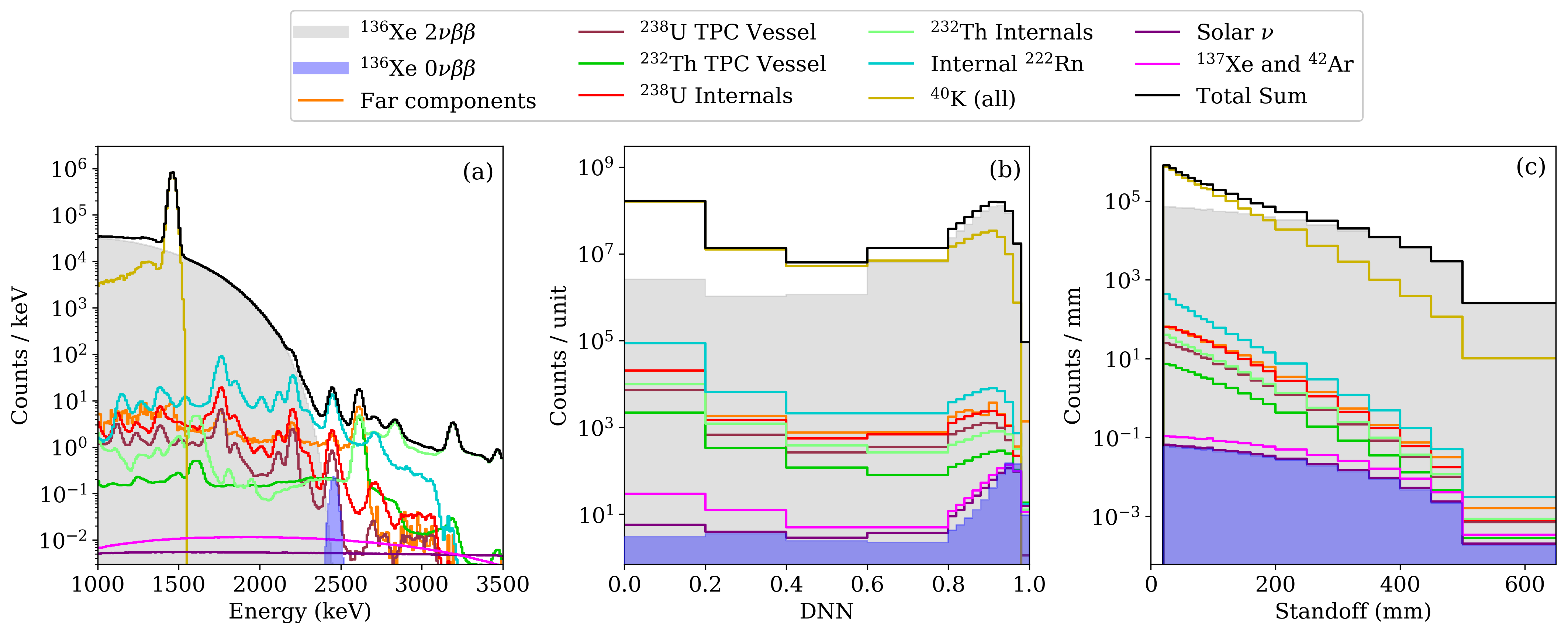}
    \caption{Nominal model of event distributions in nEXO, projected onto each of the three axes used in the sensitivity analysis: (a) event energy, (b) DNN \0 discriminator, and (c) standoff distance. The \0 decay signal corresponds to a half-life of $0.74\times10^{28}$ yr. }
    \label{fig:sensitivity_distributions}
\end{figure*}

\begin{figure*}[tp]
    \centering
    \includegraphics[width=\textwidth]{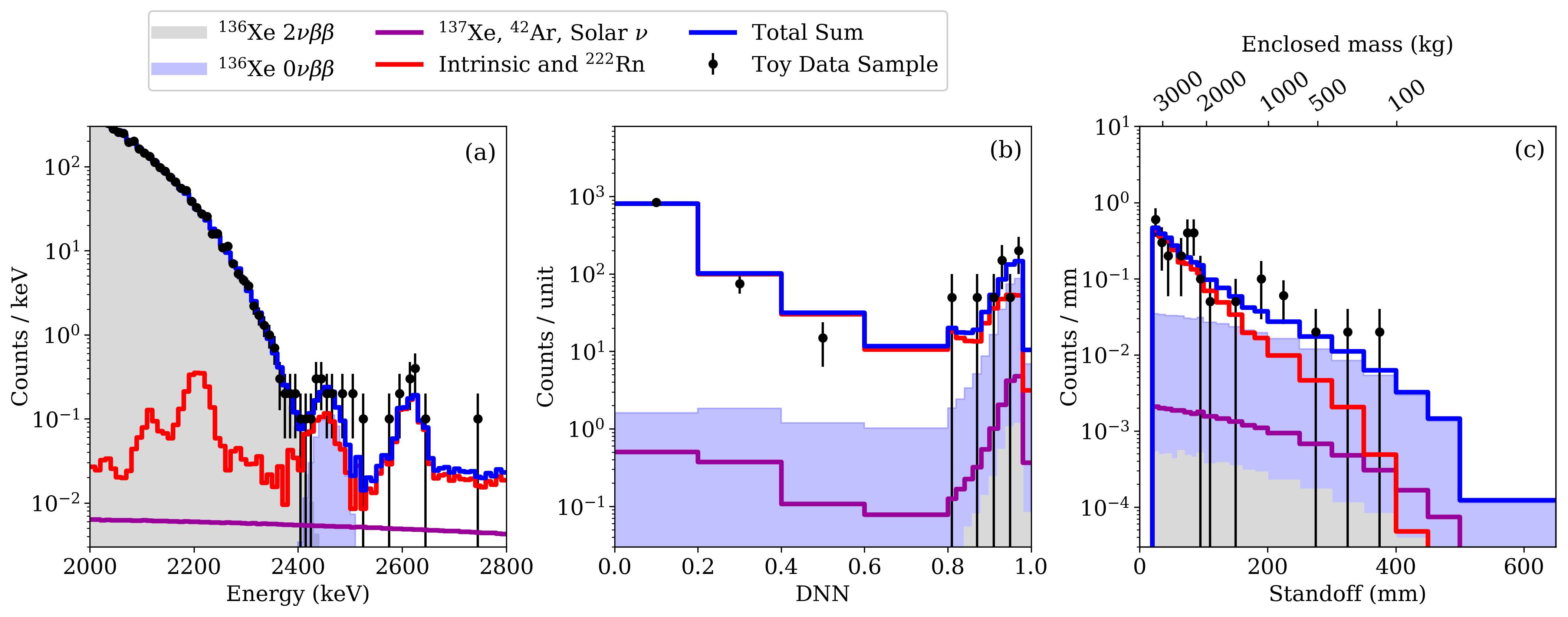}
    \caption{Event distributions for an example of toy dataset (black points) and combined groups of the fitted PDFs projected onto the three axes used in the sensitivity analysis. In (a) the event energy distribution is shown for SS-like events (DNN $>0.85$) in the central 2000~kg LXe and in the 2000-2800 keV region; (b) the DNN \0 discriminator distribution is shown for events with energy within \Q $\pm$ FWHM/2 and in the same central volume; and (c) the standoff distance distribution is shown for SS-like events within \Q $\pm$ FWHM/2. The \0 decay signal corresponds to a half-life of $0.74\times10^{28}$ yr.}
    \label{fig:standoff_in_roi}
\end{figure*}

The relative contribution of each component to its group PDF is defined by Eq.~\ref{eq:expected_counts}, and the total contribution of that group to the toy event is defined as the sum of the expected counts from each individual component. 
The full toy model is defined as a distribution where the number of events in the $j$-th bin of the three-dimensional fit space is given by the expression
\begin{equation}
    n_j = N_{0\nu} f_{0\nu\,j} + \sum_i N_i\,f_{i j}
    \label{eq:bin_values}
\end{equation}
where $N_i$ represents the expected number of events from the $i$-th group and $f_{i\,j}$ is the value of the PDF for the $i$-th group in bin $j$. The $N_i$s are variable parameters -- when building a model to generate toy datasets they are fixed at the values computed using Eq.~\ref{eq:expected_counts}, but they are allowed to float when fitting the model to a toy experiment. The projections of the model onto each of the three axes, as well as of the contribution from each group, is depicted in Figure~\ref{fig:sensitivity_distributions}. 

A negative log-likelihood (NLL) is minimized to determine the best fit values for $N_i$ in each toy dataset. The profile likelihood-ratio is used as a test statistic to build confidence intervals for exclusion or discovery of \0 decay. The details of the statistical analysis employed in this work are provided in \ref{sec:appendix}. 

An example of a toy dataset assuming that \0 decay exists and has a half-life of $0.74\times10^{28}$~yr is presented in Figure~\ref{fig:standoff_in_roi}. The figure shows the randomized data projected onto the three fitting variables along with the best fitted PDFs, which are grouped together based on similar characteristics. 
Figure~\ref{fig:standoff_in_roi} (a) shows the energy distribution of SS-like events in the central 2000~kg of LXe, where a comparable rate is observed between signal and all background components added together in the region near \Q{}. In Figure~\ref{fig:standoff_in_roi} (b), the DNN score distribution is shown for events with energy within the FWHM around \Q{} and in the central 2000~kg of LXe, while (c) shows the standoff distance distribution for SS-like events within the same energy range. There is a region where the signal is dominant over backgrounds in the bins towards \0-like DNN score and innermost volumes in the standoff distance, illustrating the \0 decay separation power obtained with nEXO.

\subsection{Sensitivity results}
\label{sec:results}

\begin{figure}[tbp]
    \centering
    \includegraphics[width=0.45\textwidth]{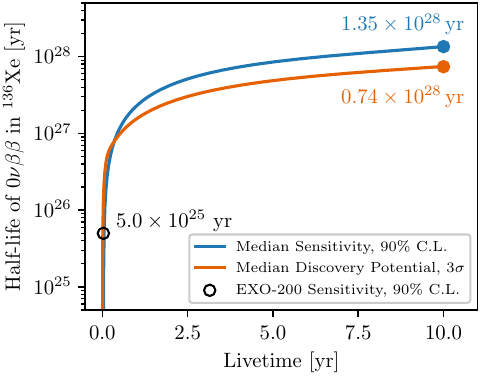}
    \caption{Projection of the median sensitivity and 3$\sigma$ discovery potential to \0 decay with nEXO as functions of the detector livetime. At small livetimes, the experiment is essentially background-free. In this regime, the number of observed counts required to make a 3$\sigma$ discovery is smaller than the number of counts that can be excluded at the 90\% CL, leading to a discovery potential that is higher than the sensitivity.}
    \label{fig:sens_vs_livetime}
\end{figure}

\begin{figure}[tbp]
    \centering
    \includegraphics[width=0.45\textwidth]{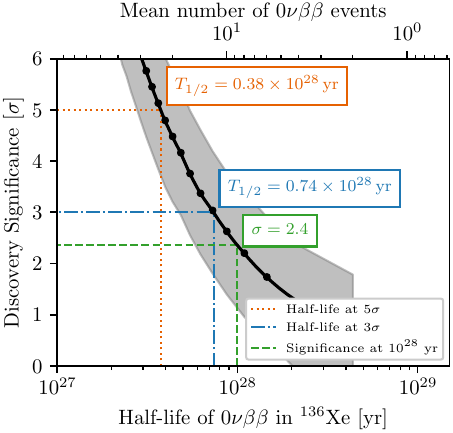}
    \caption{Projections of the median discovery significance to \0 decay with 10 years of nEXO data. The grey band indicates the 68\% symmetric quantile around the median.}
    \label{fig:dp_sigma_10yr}
\end{figure}

Figure~\ref{fig:sens_vs_livetime} shows the projected \0 sensitivity and discovery potential at 3$\sigma$ level as functions of livetime. If no signal is observed in a 10~year exposure, the median expected exclusion for nEXO is $T_{1/2}^{0\nu} > 1.35\times10^{28}$~yr at the 90\% CL. With one year of data taking this value is $0.2\times10^{28}$~yr and exceeds $10^{28}$~yr after 6.5 years. 
For a potential signal, the median $3\sigma$ discovery potential is $T_{1/2}^{0\nu} = 0.74\times10^{28}$~yr. These results represent an increase of $\sim 30-45\%$ over the previous projections reported in Ref.~\cite{PhysRevC.97.065503}. About $1/3$ of the improvement arises from the DNN discrimination power, consistent with results reported by EXO-200~\cite{PhysRevLett.123.161802}, and the remaining $2/3$ arises from the reduction of backgrounds, dominated by the choice to use EF copper. 
The expected discovery significance as a function of the \0 decay half-life at the nominal 10 year livetime is presented in Figure~\ref{fig:dp_sigma_10yr}. We observe that a median significance of $2.4\sigma$ is achieved for a hypothetical \0 decay half-life of $10^{28}$~yr. If \0 decay exists with half-life of $<0.38\times 10^{28}$~yr then nEXO would be capable of discovery at $5\sigma$ significance in more than 50\% of the possible realizations.

\begin{figure}[tbp]
    \centering
    \includegraphics[width=0.45\textwidth]{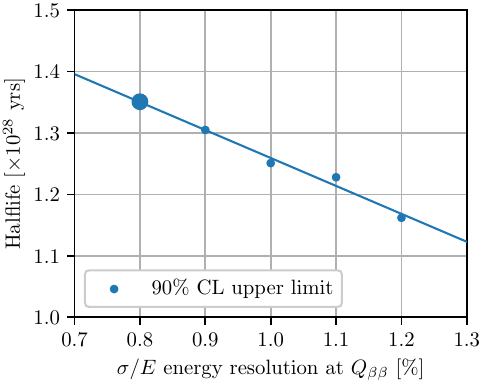}
    \caption{Projected median sensitivity to \0 decay at 90\% CL with nEXO operating at different energy resolutions. The large marker represents the value at the baseline model, while smaller points represent the calculations with the energy resolution intentionally degraded. The solid line is a linear fit to the calculated points.}
    \label{fig:sens_vs_eneresol}
\end{figure}

The energy resolution obtained in the MC simulations is based on a realistic representation of the detector response but  also includes some simplifications, a few of which cannot yet be fully validated at this stage of the experiment design. To quantify the impact of the energy resolution on the experimental sensitivity, we degraded the resolution value at \Q{} within a conservative range of values and then re-evaluated the sensitivity. The result is shown in Figure~\ref{fig:sens_vs_eneresol}. As previously reported ~\cite{PhysRevC.97.065503}, the \0 sensitivity does not degrade substantially, decreasing by $\sim 15\%$ for the resolution growing by 50\%. 

\begin{figure*}
    \centering
    \includegraphics[width=\textwidth]{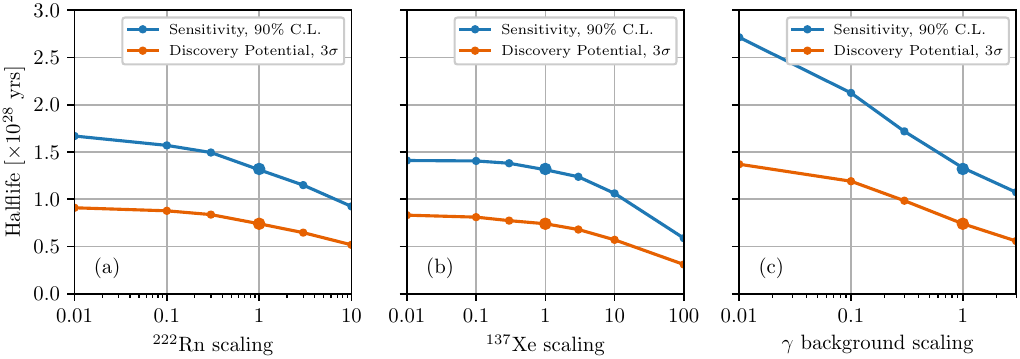}
    \caption{nEXO median sensitivity and 3$\sigma$ discovery potential evaluated for hypothetical variations in background scale from (a) \isotope{Rn}{222} decays in the LXe, (b) \isotope{Xe}{137} production, and (c) all $\gamma$ background components.}
    \label{fig:sens_vs_backgrounds}
\end{figure*}

\isotope{Rn}{222} decays in the LXe contribute approximately 50\% of the total background budget, as discussed in section~\ref{sec:materials}. Because of the complexity and inherent uncertainty of estimating the nEXO radon content by up-scaling the EXO-200 experience, R\&D, at this time focusing on devices to remove xenon contaminants, is ongoing. In particular, online removal could allow even lower levels of radon contamination thus improving the nEXO sensitivity. Therefore, the \0 sensitivity and 3$\sigma$ discovery potential are also evaluated for values between 100x smaller than the Rn-background design value and  up to 10x larger than the design value. The results are shown in Figure~\ref{fig:sens_vs_backgrounds} (a). In the most pessimistic scenario considered, the sensitivity reach would reduce to just below $10^{28}$~yr. 

Albeit subdominant, backgrounds from \isotope{Xe}{137} are particularly difficult to reject through their event topology because they arise from $\beta$ decays nearly uniformly distributed in the FV.  
Estimates of cosmogenically-produced \isotope{Xe}{137} are well understood but more advanced veto techniques than the simple scheme presented in section~\ref{sec:exposure-based-bkg} could further reduce this component. On the other hand, less is known about the \isotope{Xe}{137} production rate from neutrons from $(\alpha, n)$ reactions or spontaneous fission. Figure~\ref{fig:sens_vs_backgrounds} (b) presents the effect of a possible reduction or increase of \isotope{Xe}{137} on the median \0 sensitivity and discovery potential at 3$\sigma$, which are found to remain within $\sim 10\%$ of the baseline value up to a fourfold increase in \isotope{Xe}{137} rates. 

During further development of the experiment continued attention will be given to the benefits arising from reduced background and improved detector performance. A broader investigation of the impact from variations in the rate of $\gamma$ backgrounds was therefore performed by scaling all $\gamma$ background components (including \isotope{Bi}{214} from out-of-equilibrium \isotope{Rn}{222}) and re-evaluating the \0 sensitivity and discovery potential. The results are shown in Figure~\ref{fig:sens_vs_backgrounds}~(c). In particular, a power law fit to these curves, when plotted as a function of background index $B$, yields $T_{1/2}^{0\nu} \propto B^{-0.22}$ and $T_{1/2}^{0\nu} \propto B^{-0.26}$ for the sensitivity and   discovery potential at 3$\sigma$, respectively. The fitted power for the sensitivity represents a substantial improvement compared to the value of $-0.35$  reported in Ref.~\cite{PhysRevC.97.065503}, and brings nEXO closer to the background-free regime, where this power is zero.

To appreciate how close nEXO is to be a background-free experiment, we also considered the ideal scenario --- wherein all backgrounds are perfectly removed with the exception of \2 --- to determine the ultimate sensitivity of a LXe TPC experiment of nEXO's size. In this case, even the remaining contributions from \2 decays near \Q{} are highly suppressed by the good energy resolution.
The resulting upper limit at 90\% CL obtained with the approach employed in this work is 1.9~counts, which is smaller than 2.44 counts as expected in the optimal case of a truly zero background experiment~\cite{Feldman:1997qc}. Following the approach suggested by the same reference, we used the latter value to calculate the sensitivity reach, which in this case results in $3.6 \times 10^{28}$~yr.

\subsection{Comparison to a counting experiment}

By incorporating all aspects of the experiment, we consider the \0 decay half-life sensitivity the most representative figure-of-merit of the experiment's design and performance. For comparison purposes, it is useful to summarize the complex background spatial distribution and the performance of the multi-parameter fit into a single parameter, as would be the case for a counting experiment. 

Compared to our multi-dimensional likelihood analysis, in a counting-experiment analysis, all the information is collapsed into a single bin which only contains events passing all event selection cuts. This corresponds to SS-like events with energy within the range $Q_{\beta\beta} \pm \mathrm{FWHM}/2$ and position within the FV. The fraction of \0 events in this more narrowly-defined region of interest is 52.0\%. We assume that both the signal and background distributions are given by independent Poisson distributions with means $s$ and $b$, respectively. 

Following the procedure above, one would generate many toy experiments producing different values for the observed number of events $n$ from a Poisson distribution with a mean of either $s+b$, to evaluate the discovery potential, or simply $b$, to obtain the sensitivity reach, and then compute the $p$-value with respect to the null hypothesis. The median significance obtained from the distribution of $p$-values exhibits a counter-intuitive saw-tooth behavior as a function of background $b$ due to the discrete nature of the observable counts. Instead, we follow the procedure outlined in Ref.~\cite{Bhattiprolu:2020mwi}, where the $p$-value of the average $n$ over all toy experiments is used to calculate the significance. The use of this so-called Asimov dataset overcomes the above mentioned shortcomings and results in a monotonic dependence of the $p$-value as a function of background. 

The above procedure is used to calculate an effective background index for which such a counting experiment would yield the same half-life sensitivity at 90\% CL as the multi-dimensional likelihood analysis presented in the previous section. For nEXO's projected half-life sensitivity of \SI{1.35e28}{\year} at 90\% CL, the effective background index is equal to $b=\SI[per-mode=repeated-symbol]{7e-5}{\cts / (\fwhm \cdot \kilo\gram \cdot \year)}$. While we have explored alternative methods for calculating the sensitivity \cite{Kumar:2015tna} and discovery potential \cite{Cowan:2010js} in the case of a counting experiment, we found the above method in combination with the Asimov dataset to be the most straightforward and to be neither overly optimistic nor conservative.

\section{Conclusions}\label{sec:conclusion} 

\begin{figure}[tbp]
    \centering
    \includegraphics[width=0.99\columnwidth]{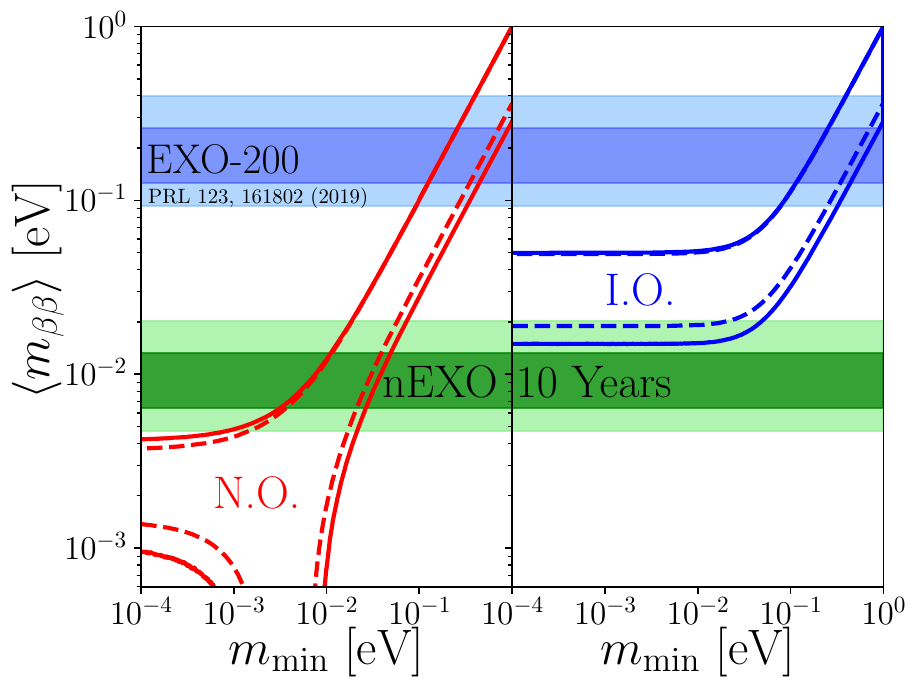}
    \caption{90\% CL exclusion sensitivity reach to the effective Majorana neutrino mass $\langle m_{\beta\beta} \rangle$ as a function of the lightest neutrino mass for normal (left) and inverted (right) neutrino mass ordering. The width of the horizontal bands derives from the uncertainty in nuclear matrix elements (see text) and the values of $\langle m_{\beta\beta} \rangle$ are calculated using $g_A = 1.27$. The darker bands covers $\sim68\%$ of the NME values. The dashed contours of the normal neutrino mass ordering (NO) and inverted ordering (IO) bands result from the unknown Majorana phases and are, to this date, unconstrained. The outer solid lines incorporate the 90\% CL errors of the three-flavor neutrino fit of Ref.~\cite{Esteban:2020cvm}}
    \label{fig:lobster}
\end{figure}

nEXO has been designed to exploit the advantages of the LXe TPC technology to search for \0 decay at the tonne scale. In this report we present an updated projection for the nEXO median sensitivity to the \0 decay half-life of \isotope{Xe}{136}, resulting in a new baseline-value of $1.35\times10^{28}$~yr at 90\% CL in 10 years of livetime. The significant improvement obtained for the sensitivity reach compared to previous reports by the Collaboration results from the success of a comprehensive R\&D program along with a refined design of the detector and improved data analysis tools.

\begin{figure}[tbp]
    \centering
    \includegraphics[width=0.45\textwidth]{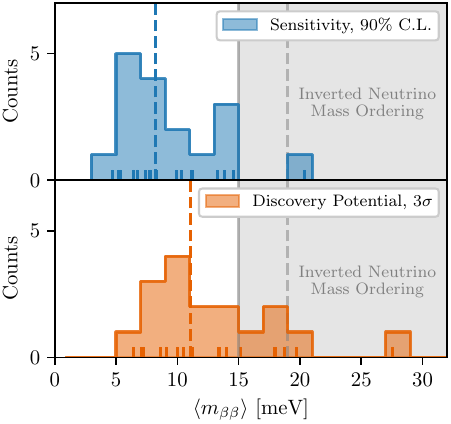}
    \caption{Effective Majorana mass $\langle m_{\beta\beta} \rangle$ sensitivity (top) and discovery potential (bottom) for nEXO for different NMEs, the standard light neutrino exchange mechanism and $g_A = 1.27$. While the $\langle m_{\beta\beta} \rangle$ is not a statistical variable, to guide the eye we also provide the medians of the $\langle m_{\beta\beta} \rangle$ ``distributions'' (vertical dashed lines of the same color). The individual  $\langle m_{\beta\beta} \rangle$ values are indicated by the tick marks on the horizontal axis. The grey area represents the inverted neutrino mass ordering band with the solid and dashed lines corresponding to the lowest possible values of the contours in Figure~\ref{fig:lobster}.}
    \label{fig:mbb_hist}
\end{figure}

Under the assumption that the \0 decay process is mediated by the exchange of a light Majorana neutrino~\cite{Avignone:2007fu}, the sensitivity to its half-life can be converted into a coverage region for the effective Majorana neutrino mass $\langle m_{\beta\beta}\rangle$ using: 
\begin{equation}
[\T]^{-1} = \frac{\mbb^2}{m^2_e}G_{0\nu}g_A^4|M^{0\nu}|^2 ,
\label{eq:T-m}
\end{equation}
where $m_e$ is the electron mass, $g_A$ the axial-vector coupling constant, 
$G_{0\nu}$ the kinematic phase-space factor, and $M^{0\nu}$ the nuclear matrix element. 
$\langle m_{\beta\beta}\rangle$ can be expressed in terms of 
the neutrino mass eigenvalues ($m_i$) and the PMNS neutrino mixing matrix elements ($U_{ei}$) by:
\begin{equation}
\mbb = \left| \sum_{i=1}^{3} U^2_{ei}m_i \right|\;.
\label{eq:mee}
\end{equation}
Figure~\ref{fig:lobster} shows the nEXO exclusion sensitivity to $\langle m_{\beta\beta} \rangle$ as a function of the lightest neutrino mass. These values are calculated using $g_A = 1.27$ and $G_{0\nu}$ from Ref.~\cite{Kotila:2012zza}. 
The allowed neutrino mass bands are derived from neutrino oscillation parameters in Ref.~\cite{Esteban:2020cvm}. The $\langle m_{\beta\beta} \rangle$ exclusion band between 4.7 and 20.3 meV arises from the full range of NMEs~\cite{Caurier2008,Rodriguez2010,Suhonen2010,Simkovic2013,Mustonen2013,Meroni2013,Vaquero2013,Engel2014,Hyvaerinen2015,Barea2015,Yao2015,Horoi2016,Menendez2018,Simkovic2018,Jokiniemi2018,Fang2018,Deppisch2020}, represented by a light green band with  NREDF~\cite{Vaquero2013} and Deformed-QRPA~\cite{Fang2018} at the minimum and maximum extremes, respectively. 
Here we take an agnostic viewpoint and consider all NMEs available in the literature, and remove only those superseded by a more recent publication from the same authors. 
The $\langle m_{\beta\beta} \rangle$ reach of nEXO fully covers the inverted neutrino mass ordering by extending below 15~meV for all but one of the NMEs, and explores a significant portion of the normal ordering. This is further illustrated in Figure~\ref{fig:mbb_hist}, where the distribution of  effective Majorana neutrino mass sensitivity and discovery potential are shown for the different NMEs.

\FloatBarrier

\section*{Acknowledgments}

The authors gratefully acknowledge support for nEXO from the Office of Nuclear Physics within DOE's Office of Science, and NSF in the United States; from NSERC, CFI, FRQNT, NRC, and the McDonald Institute (CFREF) in Canada; from IBS in Korea; from RFBR in Russia; and from CAS and NSFC in China. 
This work was supported in part by Laboratory Directed Research and Development (LDRD) programs at Brookhaven National Laboratory (BNL), Lawrence Livermore National Laboratory (LLNL), Oak Ridge National Laboratory (ORNL),  Pacific Northwest National Laboratory (PNNL), and SLAC National Accelerator Laboratory.

\appendix
\section{Statistical analysis}
\label{sec:appendix}

A binned-likelihood function is built from the model:
\begin{equation}
    \mathcal{L} = \left[\prod_j \frac{n_j^{d_j} e^{-n_j}}{d_j!} \right] \times \prod_k e^{ -(N_k - N_k')^2 / (2\sigma_k^2)  }
\end{equation}
where $n_j$ and $d_j$ are the expected and observed numbers of events in bin $j$, and the second product (indexed by $k$) contains two Gaussian constraints applied to two parameters in the fit. One of these parameters is the number of events from the $^{222}$Rn background component, which is expected to have its contribution measured \textit{in-situ} to within 10\% or better as obtained by the EXO-200 collaboration~\cite{PhysRevC.92.015503}, while the other constraint is applied to the number of events from \isotope{B}{8} solar neutrinos, because its rate is tightly constrained by the Super-Kamiokande-IV experiment~\cite{Abe:2016nxk}. The NLL is minimized for each toy dataset using the Minuit software package~\cite{1975CoPhC..10..343J,iminuit}. 

\begin{figure}[tbp]
    \centering
    \includegraphics[width=0.5\textwidth]{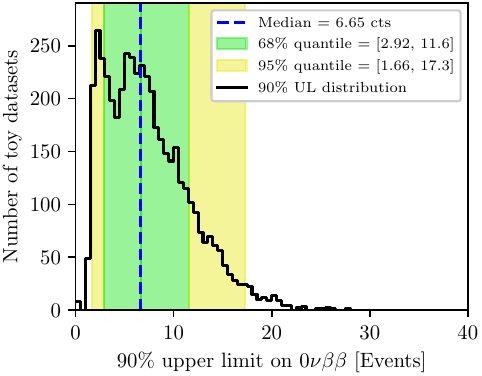}
    \caption{Distribution of upper limits $\mu_{90}$ for 5000 toy datasets. The median corresponds to a bound on the $0\nu\beta\beta$ decay halflife of  $T_{1/2} > 1.35\times10^{28}$ years.}
    \label{fig:90cl_distribution}
\end{figure}

The confidence interval to reject the \0 decay hypothesis for each toy dataset is built using the profile likelihood ratio test: 
\begin{equation}
    \lambda (\mu) = 
    \begin{cases}
    -2\,\ln\,\frac{ \mathcal{L}_\mu}{\mathcal{L}_{\hat{\mu}}} & $\hat{\mu} \geq 0$ \\
    -2\,\ln\,\frac{ \mathcal{L}_\mu}{\mathcal{L}_0} & $\hat{\mu} < 0$
    \end{cases}
\end{equation}
where $\mathcal{L}_\mu$ denotes the likelihood evaluated at a given value $\mu$ of the signal hypothesis, \emph{i.e.}, the value of $N_{0\nu}$ in Eq.~\ref{eq:bin_values}, and the single hat denotes its global best fit value. In particular, $\mathcal{L}_0$ is the likelihood for null signal, $\mu=0$. For each toy experiment, we fit a parabola to $\lambda(\mu)$ calculated for various $\mu$ hypotheses in order to determine the value $\mu_{90}$ for which hypotheses equal or larger are excluded at the 90\% or higher CL This percentile is determined by the critical value $\lambda_c(\mu)$ calculated as explained below. The distribution of $\mu_{90}$ for a sample toy dataset is shown in Figure~\ref{fig:90cl_distribution}. 

The nearly background-free nature of the nEXO experiment may invalidate Wilks's approximation for the critical value $\lambda_c$ of the NLL ratio test statistic that determines the percentiles of its distribution~\cite{Wilks1938}. For this reason, we numerically computed $\lambda_c(\mu)$ at several values of signal hypotheses $\mu$ using a frequentist approach that employs MC simulation to generate and fit an ensemble of many toy experiments as suggested in Ref.~\cite{Feldman:1997qc}. A third-order spline is used to interpolate $\lambda_c(\mu)$ for any value of $\mu$. This procedure was repeated for the different levels of backgrounds studied in this work, such as varying the detector livetime, and we found that it recovers the Wilk’s approximation $\lambda_c \simeq 2.71$ in the limit of large background rates. 

\begin{figure}[tbp]
    \centering
    \includegraphics[width=0.45\textwidth]{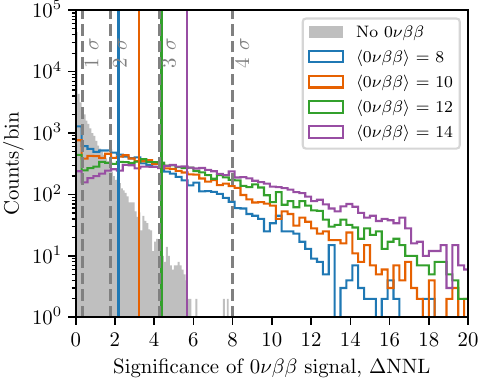}
    \caption{Sample distributions of $q_0$ for the null hypothesis $H_0$ in grey and for alternative hypotheses $H_\mu$ in color, with $\mu = 8, 10, 12$ and 14. In total \SI{1e5}{} and \SI{5e3}{} toy datasets were simulated for the null hypothesis and the alternative hypotheses, respectively. The $p$-value is calculated from counting the number of toy datasets in the distribution for $H_0$ that are above the median value for $H_\mu$, shown as vertical colored lines. The grey vertical dashed lines indicate the quantiles corresponding to $1\sigma,\, 2\sigma,\, 3\sigma$ and $4\sigma$ significance of discovery.}
    \label{fig:dp_nnl_distribution}
\end{figure}

The discovery potential to \0 decay is determined from the signal hypothesis $\mu$, $H_\mu$, which produces an ensemble of toy experiments consistent with a given median significance to reject the null hypothesis, $H_0$. We apply the following test statistic to determine the discovery significance: 
\begin{equation}
q_0 = \begin{cases}
-2\,\ln \left(  \frac{\mathcal{L}_{0}}{\mathcal{L}_{\hat{\mu}}} \right) = -2 \Delta \text{NLL} & $\hat{\mu} > 0$ \\
0 & $\hat{\mu} \leq 0$
\end{cases}
\end{equation}
where $q_0$ is null for $\hat{\mu} \leq 0$ to avoid that positive fluctuations in background counts are considered as evidence against $H_0$. This approach requires first building the distribution of $q_0$ for $H_0$, and then evaluating similar distributions for various $H_\mu$. The median of each of the latter distributions is compared to the percentiles obtained with $H_0$, and then the matching percentile is converted into a significance following Gaussian $z$-scores. For $z \geq 4$, the $q_0$ distribution for $H_0$ is approximated by a $\chi^2$ distribution because of the low statistics with these datasets. A linear interpolation is used to obtain the significance for any value of $\mu$. An example distribution of significance for $H_0$ and $H_\mu$, for $\mu = 8, 10, 12$ and 14 is shown in Figure~\ref{fig:dp_nnl_distribution}.

\section*{References}
\bibliographystyle{iopart-num}
\bibliography{refs_clean_newnmes.bib}

\end{document}